# Advancements and Challenges in Sintering $Cr^{4+}$:YAG Ceramics: A Comprehensive Review on Transparent Ceramic Technology for Q-switched Lasers.

Mykhailo Chaika*.

*Institute of Low temperature and Structure Research, Polish Academy of Sciences, Wroclaw, 50 – 422, Poland*

\* Corresponding author.

E-mail: m.chaika@intibs.pl

**Abstract:** The resent progress in the technology of transparent ceramics extends the application of CW and pulsed lasers. The parameters of the transparent ceramics are comparable with the single crystals both for active elements (Nd:YAG, Yb:YAG, etc.) and passive Q-switchers such as $Cr^{4+}$:YAG. Despite the reaching of considerable progress, the sintering of $Cr^{4+}$:YAG remains a challenge up to now. Up to the present the influence of the CaO, MgO and $Cr_2O_3$ additives on the sintering process is poorly understander. Moreover, there is an absence of the unified model of $Cr^{4+}$ ions formations in YAG ceramics. This factor limits the progress in the development of $Cr^{4+}$:YAG ceramics. This review focused on the influence of the single additives and their combinations on the sintering trajectory of tetravalent chromium - doped YAG ceramics and the $Cr^{4+}$ ions formation.

**Keywords:** Transparent ceramics, $Cr^{4+}$:YAG, Ceramic technology, Advanced ceramics, Sintering aids.

# 1 Introduction

Over the past decade, solid-state lasers have been widely used in many fields, such as: metalworking, medical applications, light sources in laser printers and projectors, barcode readers, video players, environmental monitors, optical transmission systems, etc. [1]. Solid-state laser, in its simplest form, consists of three elements: a solid-state laser material with smooth and parallel polished faces, a system of mirrors that form a resonator, and pump sources. To start the generation, it is necessary that the gain in the laser exceeds the loss in the resonator. These losses are the sum of losses in the active medium (scattering and/or absorption of light on structural defects) and losses on mirrors.

The general trend in solid-state lasers technology is related to the increasing demand for pulsed lasers. The main parts of a pulsed solid-state laser are an active element and a Q-switcher. There are two types of Q-switched lasers: active and passive. The main advantage of active Q-switched lasers is the possibility to moderate the parameters of the laser pulse, while in the other case they require additional equipment for controlling their parameters. In the other hand passive Q-switched lasers require only presence of saturable absorber inside of resonator, and therefore is popular in production compact pulsed lasers (Fig. 1.1). The most popular host for solid-state lasers is yttrium aluminum garnet (YAG) due to its excellent mechanical and thermophysical properties. Nd(Yb):YAG and $Cr^{4+}$:YAG are used as active elements and Q-switcher, respectively. Particular attention should be paid to $Cr^{4+}$:YAG, because along with the advantages - the ability to receive sub-nanosecond pulses of high power and durability, there are disadvantages - significant heat dissipation because 85% of absorbed energy is released as heat. Despite such disadvantages in the properties, $Cr^{4+}$:YAG ceramics remains nonalternative material for passive Q-switched laser device operate at 1 μm.

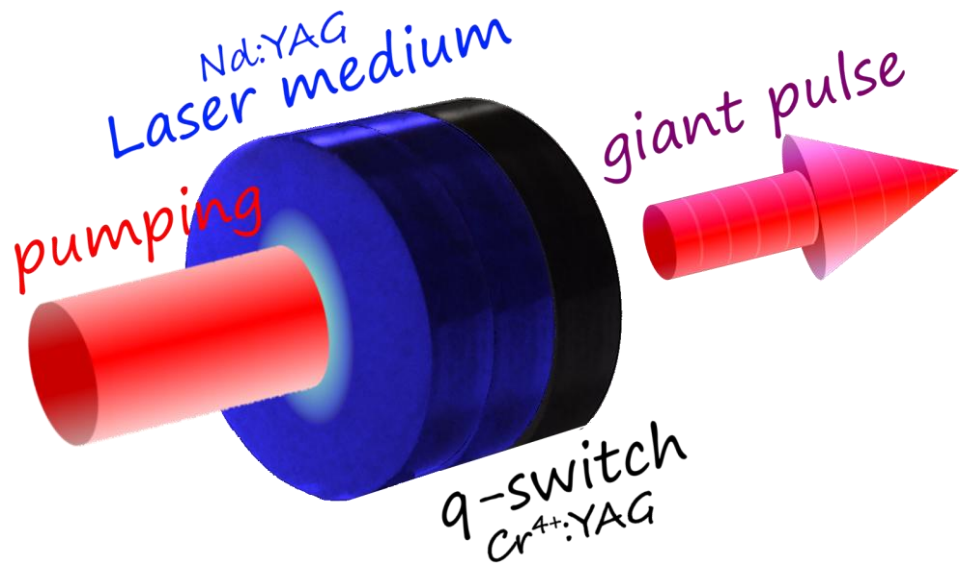

Fig. 1.1: Passive Q-switched laser element

In the last decade, the technology of the manufacture of laser-quality Nd(Yb):YAG ceramics has achieved significant success and has several advantages over technologies for growing similar single crystals both from an economic point of view and from technological flexibility in the production of ceramics in different shapes and sizes [1–3]. Today, the lasers based on Nd:YAG ceramics have reached a continuous generation power the order of megawatts. However, several problems arise on the production of $Cr^{4+}$:YAG ceramics laser quality.

Solid state reaction (SSR) sintering in vacuum the one of the popular methods for synthesis of transparent YAG ceramics laser quality using $Y_2O_3$, $Al_2O_3$ powders and additives-modifiers. However, to obtain $Cr^{4+}$:YAG ceramics, additional functional additives are required to compensate for the difference between the charge state of $Cr^{4+}$ and $Al^{3+}$ ions, since that the tetravalent chromium replaces the trivalent aluminum ion in the garnet crystal lattice. These additives can affect a sintering trajectory and, as result, the morphological and optical properties of the final ceramics. The key to manufacturing high quality $Cr^{4+}$:YAG ceramics lies in the control of the sintering parameters and, in, the kind and concentration of additives.

The main motivation of publishing this article is collecting all information related to the effect of additives-modifiers on the structure and charge state of chromium ions in $Cr^{4+}$:YAG ceramics made by SSR sintering. The better understanding of the processes which occur during SSR sintering will allow us to increase the properties of Cr-doped YAG ceramics in the future. It should be noted that the properties of synthesized ceramics depend on the type and/or concentration of dopants and from the whole procedure of ceramics preparation. For example, the change in the heating speed during vacuum sintering can change the interaction between the dopants which will affect the final properties of ceramics. The same additives in the same concentrations will have different effects on the properties of ceramics reported by the different authors. For synthesized high-quality Cr:YAG ceramics, the improvement of the sintering process is necessary. For example, the change of speed of planetary mills and the number of balls will have a bigger effect on the transparency than change of the concentrations of functional additives. This paper will allow us to understand the role of CaO, MgO, and $Cr_2O_3$ additives on the sintering trajectory and find proper parameters to synthesized high quality $Cr^{4+}$:YAG ceramics. It should be noted that a major fraction of conclusions is based on the results reported. Therefore, the proposed explanations might be affected by the lack of data and differ from the actual cases.

## 2 Sintering of $Cr^{4+}$:YAG ceramics

Both changes in the concentrations of additives and the sintering parameters influenced the sintering route of YAG ceramics. The role of the sintering parameters on the ceramic's fabrication were earlier published by many authors in different papers, books, and review [1,4–6]. The same processes occur in the case of $Cr^{4+}$:YAG ceramics and the described mechanism of evolution of microstructure will be the same. However, the difference is that the sintering of $Cr^{4+}$:YAG ceramics requires presence of $Ca^{2+}$ or/and $Mg^{2+}$ for stabilize the chromium ions in tetravalence state [7–10]. The presence of these additives changes the sintering trajectory of Cr$^{4+}$:YAG ceramics. This chapter describes the feature of sintering of Cr$^{4+}$:YAG ceramics.

## 2.1 The influence of SiO$_2$ additive on the sintering of Cr$^{4+}$:YAG ceramics

Sintering of Re/Tm-doped YAG transparent ceramics achieved high progress in the last decades, but there are many unresolved issues in the case of Cr$^{4+}$:YAG ceramics. Kuklja et al. proposed that the incorporation of Si$^{4+}$ in the YAG crystal lattice introduces an extra positive charge [11]. To compensate for this charge, the cation vacancies are generated. The appearance of cation vacancies increases Y$^{3+}$ and Al$^{3+}$ ions diffusivity in the YAG, which cause an increase in ceramics densification compared to Si-undoped YAG [12]. TEOS sintering additive works also for the case of sintering ceramics directly from YAG powders [13] proven the vacancies formation theory.

The feature of Cr$^{4+}$:YAG materials is the necessity for doping with divalent dopant, such as Ca$^{2+}$ or Mg$^{2+}$ ions to stabilize the Cr$^{4+}$ ions. Firstly, Cr$^{4+}$:YAG ceramic was made in 1996 by A. Ikesue [14] and Ca$^{2+}$ and Mg$^{2+}$ ions were used as a charge compensator together with TEOS as sintering aid. However, the sintered ceramics quality was lower than that of single-crystal analogues. The next attempt to make Cr$^{4+}$:YAG ceramics was made in 2006 [15]. The sintered ceramics had a good concentration of tetravalent chromium ions; however, a further development had not come until 2015 [16,17]. Since that, the number of papers was grown, but until now many questions are still unclear [10,16–27].

The main characteristic of Cr$^{4+}$:YAG materials is the necessity to use divalent additives for charge compensation. Yttrium-aluminum-garnet crystal (YAG) has a cubic structure and belongs to the Ia-3d space group with the stoichiometric formula C$_3$A$_2$D$_3$O$_{12}$, where C, A and D denote dodecahedral, octahedral, and tetrahedral lattice sites, respectively (Fig. 1.2) [28,29]. The possibility of using Cr$^{4+}$:YAG materials as Q-switched lasers is based on the large ground-state absorption cross-section and excited-state lifetime of tetrahedral coordinated Cr$^{4+}$ ions.

During vacuum sintering, Cr dopants incorporate into the YAG lattice in the trivalent state, replacing $Al^{3+}$ in the octahedral positions [27,30]. After vacuum sintering, only $Cr^{3+}$ exists, because the formation of $Cr^{4+}$ requires high-temperature air annealing. $Cr^{4+}$:YAG ceramics should contain divalent additives such as $Ca^{2+}$ or $Mg^{2+}$ for stabilizing $Cr^{4+}$ in the tetravalent state [24].

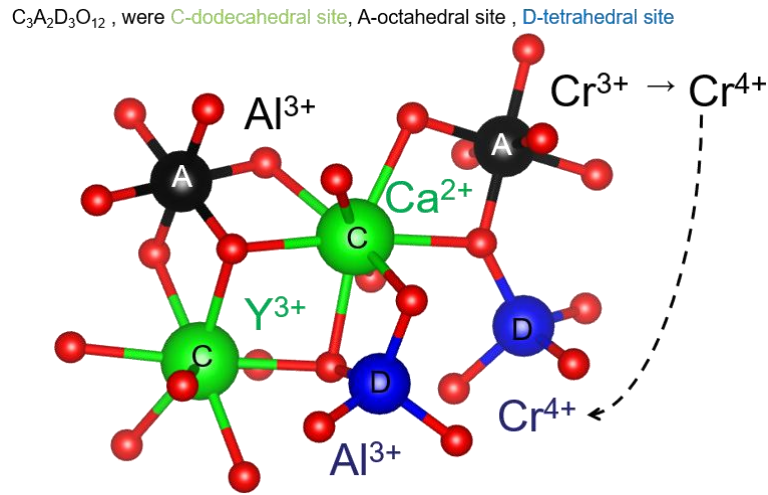

Fig. 1.2. Schematic illustration of structure, occupancy, and $Cr^{3+} \rightarrow Cr^{4+}$ ions valence transformation in $Cr^{4+}$:YAG ceramics.

One of the ways to create highly transparent YAG ceramics is using $SiO_2$ sintering aids. Ikesue et al. [2] first reported highly efficient laser gain of transparent Nd:YAG ceramics produced by sintering a powdered mixture of $Al_2O_3$, $Y_2O_3$, and $Nd_2O_3$ at 1750°C for 8 h. They found that the addition of 0.14 wt.% (1.35 mol%) $SiO_2$ is crucial for sintering transparent Nd:YAG ceramics [6]. This phenomenon has been later confirmed and, nowadays, different concentrations of $SiO_2$ are commonly added as a sintering aid to produce Re:YAG ceramics of laser quality. The most popular model for describing the interaction between $SiO_2$ and YAG is the formation of a liquid phase [31] or an increase in the concentration of cation vacancies due to the incorporation of $Si^{4+}$ ions into the YAG lattice [32,33].

The main difference of $Cr^{4+}$:YAG ceramics is that addition of TEOS sintering additives have negative influence on the concentration of $Cr^{4+}$ ions [13,16,18,34]. This is occurring due to the competition of $Cr^{4+}$ and $Si^{4+}$ ions among the charge compensation additives, namely $Ca^{2+}$ or/and $Mg^{2+}$ ions [16,34]. Moreover, in some cases addition of $Me^{2+}$ ions ($Me^{2+}$ - $Ca^{2+}$ and/or $Mg^{2+}$) and TEOS sintering additives decrease transparency of synthetized samples [18,34]. Therefore, it is preferential of sintering $Cr^{4+}$:YAG ceramics without using TEOS sintering aid.

In this section discussed the influence of the interaction of MeO and $SiO_2$ (Me -Ca, Mg) on the formation Cr,Ca:YAG ceramics. The interaction of $CaO$-$SiO_2$ at the grain boundary caused formation of different intermediate phases thus chancing sintering route of Cr,Ca:YAG ceramic. For example, some specific ratio of $CaO/SiO_2$ additives caused appearance of the liquid phase at the grain boundary resulting in abnormal grain growth. Mutual co-doping YAG ceramics with $Si^{4+}$ and $Me^{2+}$ ions caused formation of [$Me^{2+}$... $Si^{4+}$] charge neutral complex thus depleting positive influence of $Si^{4+}$ ions on ceramics densification and on the $Cr^{4+}$ ions concentration.

*2.1.1 The mutual influence of CaO and $SiO_2$ on the microstructure of $Cr^{4+}$:YAG ceramics*

Sintering of transparent ceramics depends on the properties of initial compounds prior the sintering. There are different approaches related to the source of $Ca^{2+}$ ions, for example its can be used CaO [24,34,35], $CaCO_3$ [25,36], $Ca(NO_3)_3$ [37,38], etc. as raw materials. The CaO remains the main additives towards sintering of $Cr^{4+}$:YAG ceramics. Therefore, this subsection describes the influence of interaction of CaO and $SiO_2$ on the sintering $Cr^{4+}$:YAG ceramics.

The optical properties of $Cr^{4+}$:YAG ceramics synthetized with [34] and without [27] TEOS sintering aid exhibits difference in the microstructure. It was proposed that this difference is due to the interactions between CaO and $SiO_2$ sintering aid during vacuum sintering. The interactions between these additives occurs thorough formation of calcium silicates during the heating stage, changes the sintering trajectory of $Cr^{4+}$:YAG ceramics. During the heating stage, the ceramic density increased from 70 to 99.9 %, while only a fraction of the porosity was removed during isothermal annealing [32]. SSR sintering occurs by interaction of the main compounds of $Y_2O_3$, $Al_2O_3$ thorough the formation of intermediate phases $Y_4Al_2O_9$, $YAlO_3$ and finally YAG [19,39]. Formation of garnet structure occurs at low temperature ~ 1300 °C [19]. Due to its low solubility, CaO and $SiO_2$ remains longer on the grain boundary than $Cr_2O_3$ and, therefore can interact between itself. Indirect confirmation of CaO-$SiO_2$ interaction during the synthesis of YAG ceramics can be extracted from the chemical analysis of the ceramics. It was detected that the ceramics with the TEOS sintering additive exhibits reduction of Ca concentration after vacuum sintering [34], while the opposite was found for the ceramics without TEOS [27].

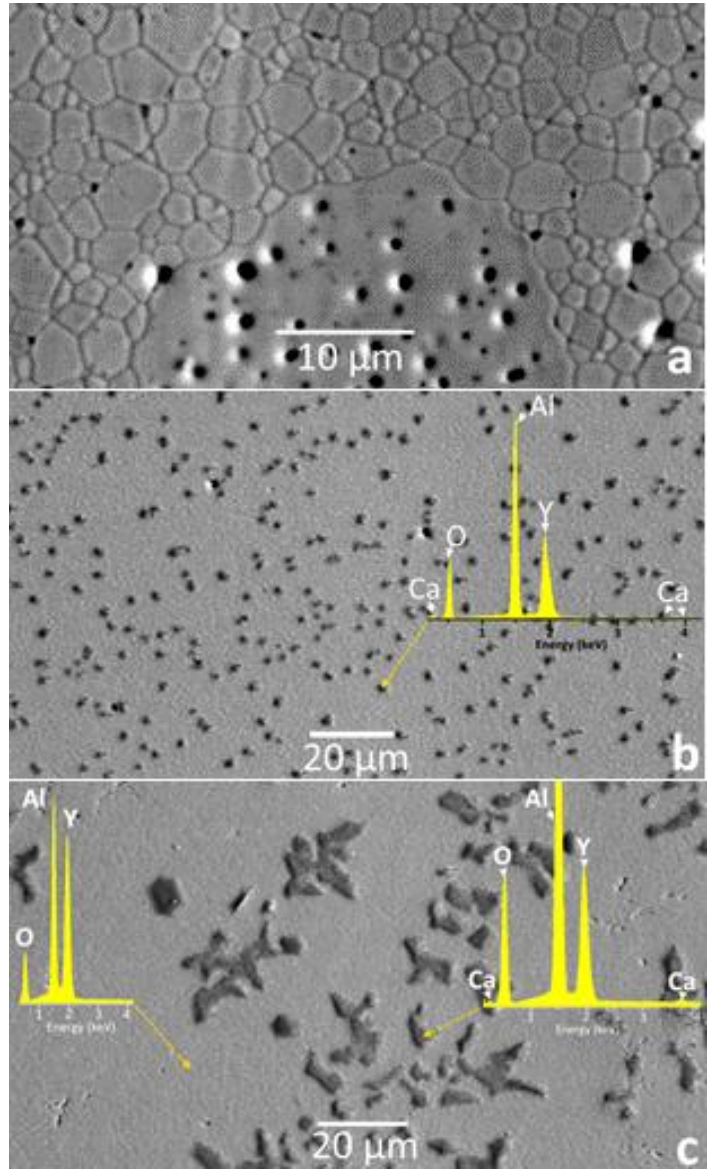

Fig. 2.1: SEM images of Si,Cr(0.1at.%),Ca(0.5at.%):YAG (b) Si,Cr(0.1at.%),Ca(0.8at.%):YAG, and (c) Si,Cr(0.1at.%),Ca(1.2at.%):YAG (d) ceramics [34].

The influence of the interaction between the CaO and $SiO_2$ depends on their ratio used for sintering of $Cr^{4+}$:YAG ceramics [18,34]. Using the same concentration of TEOS (0.5wt.%) and increase in the concentration of $Ca^{2+}$ ions from 0.5at.%, 0.8at.%, to 1.2at.% allow to change of Si/Ca ratio from ~$1/4$, respectively. For example, the Si,Cr(0.1at.%),Ca(0.5at.%):YAG ceramics (Si/Ca ratio ~$1/2$) characterized by abnormal grain pore growth with the capturing of pores inside the grains (Fig. 2.1(a)) [34]. The decrease of Si/Ca ratio to ~$1/3$, and ~$1/4$, prevent abnormal grain growth with the formation of smaller grains [34]. This drastically change in the microstructure indicates that the sintering trajectory of these ceramics depends on the concentration of divalent dopant. In fact, it was proposed that the main reason for the change in microstructure was the change of Si/Ca ratio.

Proposed explanation was in agree with the reported results for other Cr-doped YAG ceramics. The absence of abnormal grain growth was found for Cr(0.1at.%),Ca(0.5at.%):YAG [27] and for Ca(0.5at.%):YAG [40] ceramics synthesized by the same methodology but without TEOS sintering additives indicates that the abnormal grain growth for the ceramics with Si/Ca ratio ~$^1/_2$ caused by the presence of both CaO and $SiO_2$. It should be noted that the change in the sintering method will also affect the microstructure. The absence of abnormal grain growth for Ca-doped YAG ceramics was not detected in Cr:YAG [41], Cr,Ca:YAG [13], and Ca,Mg,Cr:YAG [25] ceramics. However, it also was reported that Ca,Mg,Cr:YAG ceramics at certain concentration of dopants (in contrast to the Si,Ca,Mg,Cr:YAG ceramics of the same compositions) show abnormal grain growth as in our case [16]. At the same time, the abnormal grain growth was often reported YAG ceramics us in the case of pure YAG ceramics [35,42], or for YAG ceramics with the certain amount of TEOS sintering additive [39]. This once again shown that the sintering of YAG ceramics is more complicated that described in this review.

The change $SiO_2$–CaO ratio influenced their interactions and different intermediate phases can be formed. It was proposed that during the sintering of the ceramics with Si/Ca ratio ~$^1/_2$, the interaction between $SiO_2$ and CaO might lead to formation of eutectic composition which melts at 1450 °C thus changing the sintering trajectory resulting in abnormal grain growth. Instead, the liquid phase due to the $SiO_2$–CaO interaction in the ceramics with Si/Ca ratio ~$^1/_3$, and ~$^1/_4$ can occur at the higher temperature 2057 °C. Detailed explanation can be found in our paper [34]. It should be noted that the earlier proposed explanation of the role of TEOS sintering additives is based on the interaction between $SiO_2$ and YAG at the heating stage which helps to obtain high dense ceramics [31]. Therefore, the interaction of $SiO_2$ and CaO on grain boundary eliminate this positive influence of $SiO_2$ because all silicon oxide is bounded with the calcium oxide in different intermediate phases. Moreover, the presence of these calcium silicate phases affects the microstructure transformations itself.

Formation of low temperature eutectic melts during ceramics change the stability of the pore network. SSR sintering of YAG transparent ceramics can be divided at heating, and isothermal annealing stages. During the heating stage, ceramic microstructure goes thorough the: (i) formation of necks between the grains, (ii) necks evolution into grain boundaries, (iii) formation of an open pore network, (iv) thinning of the diameter of the pore channels, (v) collapse into closed pores (Fig. 2.2). The evolution of open pore network onto the chain of the closed pores depends on the relation between mass transport on grain boundary and surface of the pores [43]. It is important to remain the open pores as small as possible, because their collapse produces the chain of small pores which can be removed more easily than larger pores. The formation of different inter-grains phases (i.e., in the case of CaO and $SiO_2$ interactions) can change the diameter of the pore channels, thus affecting the properties of the final Cr,Ca:YAG ceramics.

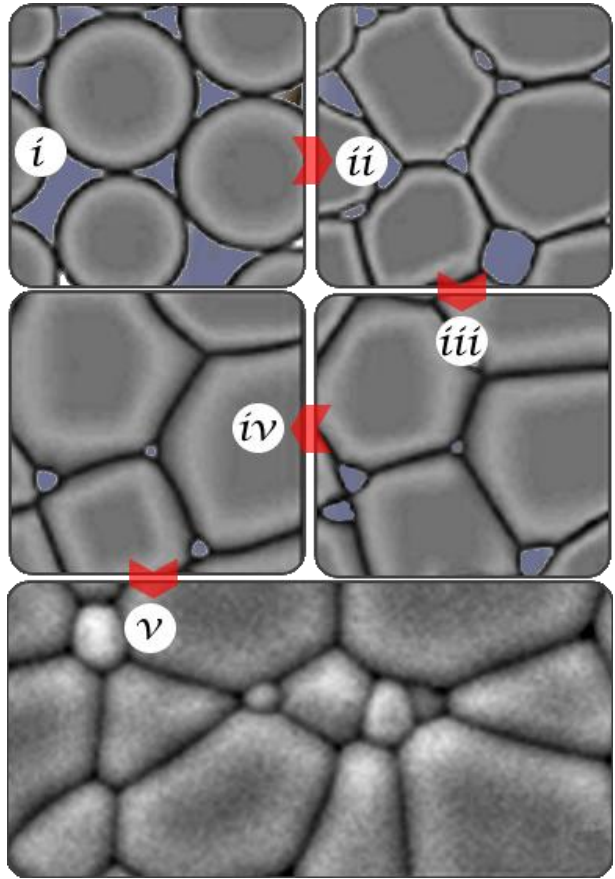

Fig. 2.2: Microstructure evolution of YAG ceramics during SSR sintering.

Co-doping $Cr^{4+}$:YAG ceramic with $Ca^{2+}$ ions and $Si^{4+}$ ions wannish the positive influence of the last on the ceramics sintering. Incorporation of $Si^{4+}$ ions in YAG lattice occurs by appearance of cation vacancies which helps to remove residual porosity thus increase the transparency of YAG ceramics. However, co-doping with the $Ca^{2+}$ ions caused formation of $[Ca^{2+}... Si^{4+}]$ charge neutral complexes without extra cation vacancies. The solubility of $Ca^{2+}$, and $Si^{4+}$ ions in YAG lattice is low and close to 300- 400 ppm [44], and up to 600 ppm [45], respectively. But the mutual solubility of these ions in significantly higher [46], which is in favor of proposed theory of formation of $[Ca^{2+}... Si^{4+}]$ charge neutral complexes. Most probably in the case of YAG ceramics, $Si^{4+}$ combines with $Ca^{2+}$ when there is enough $Ca^{2+}$, and in this case, the generation of cation vacancies does not occur. The lack of cation vacancies causes the deterioration of the ceramic densification, which results in a worsening of its optical quality [34].

The negative influence of the SiO$_2$/CaO interaction on the microstructure of Cr$^{4+}$:YAG ceramics can be by changes of synthesis route. In fact, there many papers report high quality Me,Cr:YAG ceramics with using TEOS sintering additives [13,16,23,25]. However, at the same time there is plenty works reported sintering of high quality Me,Cr:YAG ceramics without using TEOS [16,20,27,41]. However, most of authors report the negative influence of Si$^{4+}$ ions on conversional efficiency of Cr$^{4+}$ ions [13,16,23,34]. Therefore, it is preferential sintering Cr$^{4+}$:YAG ceramics without TEOS sintering additives.

*2.1.2 The mutual influence of MgO and SiO$_2$ on the microstructure of Cr$^{4+}$:YAG ceramics*

The influence of the mutual additives of MgO and SiO$_2$ on the sintering trajectory of Cr;YAG ceramics is unknown now. There is an absence of the results related to Si,Cr,Mg:YAG ceramics. However, there are plenty papers reporting the influence of mutual additives of MgO and SiO$_2$ on the optical properties of undoped and RE-doped YAG ceramics. The sintering of Cr-doped YAG ceramics differs from the sintering of RE-doped YAG ceramics [19], and therefore, it can change the interaction of MgO and SiO$_2$. But still, the main trend can be extracts from the results of sintering Cr-undoped ceramics.

One such paper reports the effect of MgO and SiO$_2$ additives on properties of undoped YAG ceramics. It was reported that the ratio 1/1 of Si$^{4+}$ to Mg$^{2+}$ ions in YAG ceramics caused a decrease an optical property in compared to the other values [47]. This result was explained by the formation of [Mg$^{2+}$...Si$^{4+}$] charge neutral complexes, like the case of Si$^{4+}$/Ca$^{2+}$ interaction. The change of Si$^{4+}$/Mg$^{2+}$ ratio caused drastically improve the transparency of obtained materials due to appearance of "free" Si$^{4+}$ or Mg$^{2+}$ ions [47].

Apart from highest MgO-doped ceramics, YAG ceramics with the MgO/SiO$_2$ ratio different from 1/1 characterized by the same optical quality [47], indicating the absence of interactions of MgO and SiO$_2$ on grain boundary. However, it was proposed also that the interaction between MgO and SiO$_2$ is possible and intermediate MgSiO$_3$ phase can be presented in TAG ceramics during ceramic sintering [48]. Moreover, the sample with MgO/SiO$_2$ ratio equal to 1/1 showed the highest optical transmittance compared with the results reported for YAG ceramics [47]. Similar result was reported also by other group were using MgO/SiO$_2$ ratio 1/1 brings best transparency for Nd:YAG ceramics [49]. It is important that these ceramics have better transparency than for ceramics with TEOS or MgO separately, which was explained by the presence of MgSiO$_3$ liquid phase which appears at the temperature of lower than 1380 °C. This indicates that two theories of MgO/SiO$_2$ interactions can be proposed: formation of intermediate phases and/or formation of [Mg$^{2+}$...Si$^{4+}$] charge neutral complexes.

The interactions of CaO with SiO$_2$, and MgO with SiO$_2$ have can be explained in the same way: formation of liquid phase at low temperature, and formation of [Me$^{2+}$...Si$^{4+}$] charge neutral complexes in YAG lattice. The main difference is the influence on the microstructure of the YAG ceramics. The interactions of MgO with SiO$_2$ could have positive effect on the optical properties, while in the case of CaO it opposite. However, the presence of SiO$_2$ in Cr$^{4+}$:YAG ceramics decrease the concentration of Cr$^{4+}$ ions which limits their use despite the positive influence on transparency as in case of MgO/SiO$_2$ doped ceramics.

*2.1.3 The mutual influence of MgO, CaO, and SiO$_2$ on the microstructure of Cr$^{4+}$:YAG ceramics*

The most used divalent additive is combination of CaO and MgO together. Using Si/Ca ratio ~$1/2$ for Si,Cr,Ca:YAG [34] or Si/Mg ratio $1/1$ for Si,Cr,Mg:YAG [47] ceramics caused drastically change the microstructure in compared to the others ratio. However, Si,Cr,Ca,Mg:YAG ceramics [16] shown absence of such extremums at the similar Si/Ca or Si/Mg ratios. It was synthesized the series of Si,Cr,Ca,Mg:YAG ceramics with the Ca/Mg ratio ~ $1/1$ and the constant amount of TEOS (0.15wt.%) with the varying of the Me (Me - Ca + Mg) concentrations, resulting in the series of ceramics with Si/Me ratio ~ $1/2$, ~ $1/4$, ~ $1/6$, and ~ $1/8$

[16]. Si,Cr,Ca,Mg:YAG ceramics with the Si/Mg ratio ~$^1/_1$ (Si/Me ratio ~ $^1/_2$), and ceramics with the Si/Ca ratio ~ $^1/_2$ (Si/Me ratio ~ $^1/_4$) characterized by the same microstructure as the other ceramics [16], which differs from the Si,Cr,Ca:YAG [34] or Si,Cr,Mg:YAG [47] ceramics, where were detected the abnormal grain growth and drop in transparency, respectively. Probably that the presence of both $Ca^{2+}$ ions and $Mg^{2+}$ ions compensate the negative effects of each other's. The increase in the Si/Me ratio caused decrease in the transparency of the synthesized samples due to the negative influence of divalent dopants, because the similar trend was found for TEOS-free ceramics synthesized in the same conditions. Nevertheless, Si,Cr,Ca,Mg:YAG ceramics had higher transparency than Cr,Ca,Mg:YAG ceramics indicates the positive influence of TEOS on the transparency.

*2.1.4 Suppression of $Cr^{4+}$ ions formation in Cr:YAG ceramics with $SiO_2$*

The presence of TEOS sintering additives decrease the concentration of $Cr^{4+}$ ions in Cr:YAG ceramics. The difference in the $Cr^{4+}$ ions concentration in sample with and without TEOS sintering additives has been reported numerous times. For example, Cr,Ca,Mg:YAG [16], and Cr,Ca:YAG ceramics [34] with TEOS sintering additives have in one order of magnitude lower concentration of $Cr^{4+}$ ions than the ceramics with the similar compositions. The reason for this was explained by the competition of $Si^{4+}$ with $Cr^{4+}$ ions on charge compensation involving $Ca^{2+}$ ions. The influence of the $Si^{4+}$ ions on the $Cr^{4+}$ ions formation can be described thorough charge compensation mechanism of $Ca^{2+}$, $Cr^{4+}$, and $Si^{4+}$ ions. It should be noted that the described model is general accepted now, however our recent discoveries indicate that the proposed explanation should be modified. More details see in chapter 3.

Sintering of $Cr^{4+}$:YAG ceramics occurs in a vacuum at ~ 1600-1800 °C and with the followed air annealing at ~ 900-1500 °C. During vacuum sintering, incorporation of $Ca^{2+}$ ions in YAG lattice occurs in the form of $[Ca^{2+}... ^1/_2V_O]$ charge neutral complex. During air annealing, oxygen from the ambient atmosphere incorporated into YAG replaces oxygen vacancy from

$[Ca^{2+}... {}^{1}/_{2}V_{O}]$ charge neutral complex thus destroying it. Therefore, $Ca^{2+}$ recharge chromium ions in tetravalent state forms $[Ca^{2+}...Cr^{4+}]$ charge neutral complex. The influence of $Si^{4+}$ ions on concentration of $Cr^{4+}$ ions is mediated by the formation of $[Ca^{2+}... Si^{4+}]$ neutral complexes, which vanish part of $Ca^{2+}$ ions necessary to compensate the $Cr^{4+}$ positive charge [40]. The presence of $Si^{4+}$ leads to the formation of charge neutral complexes $[Ca^{2+}... Si^{4+}]$ instead of $[Ca^{2+}...V_{O}]$, thus excluded parts of $Ca^{2+}$ ions from $Cr^{3+} \rightarrow Cr^{4+}$ ion valence transformations processes.

**2.2 The influence of CaO and/or MgO additives on sintering of YAG ceramics**

As concluded in the earlier section, it is preferential to exclude TEOS from sintering of $Cr^{4+}$:YAG ceramics [18,50] because $Si^{4+}$ ions substitute part of $Al^{3+}$ in the tetrahedral sites and suppress $Cr^{3+} \rightarrow Cr^{4+}$ ion valence transformation. Therefore, the high quality $Cr^{4+}$:YAG ceramics must be sintered without using TEOS. Moreover, the general trend in transparent ceramics technologies goes toward sintering aid free approaches. At the first-time synthesis of garnet transparent ceramics without sintering additives was demonstrated in 2017 [51].

Cr doped YAG ceramics originally include Cr atoms in its trivalent state, which can populate "A" sites only, because the ionic radius of $Cr^{3+}$ is too large to occupy the tetrahedrally coordinated sites (Fig. 1.2). To be used in the above-mentioned applications, part of the $Cr^{3+}$ must be oxidized to $Cr^{4+}$, which can fit into the tetrahedral "D" sites of the YAG crystal lattice, substituting part of the $Al^{3+}$ ions (Fig. 1.2). The extra charge generated in the YAG lattice after the chromium oxidation is usually compensated through the addition of co-dopants such as $Ca^{2+}$ or/and $Mg^{2+}$ [23,52]. A considerable excess of $Ca^{2+}$ ions is used for charge compensation in crystals. However, the optical properties can be significantly affected even in the case of minor variations in the concentration of functional additives used during ceramics manufacturing. For example, Ikesue et al [2,53] have found that in the case of Nd:YAG ceramics, the amount of silicon added, and the cooling rate adopted during the sample preparation change the thickness of the grain boundary phase, which affects the optical scattering loss.

The main challenge towards sintering of high-quality $Cr^{4+}$:YAG ceramics is eliminating pores from the ceramics body to achieve high transparency. Up to the present day, the main approach to achieve high transparent garnet ceramics is based on the using TEOS sintering additives. However, this approach does not work with the $Cr^{4+}$:YAG, therefore the sintering trajectory of this ceramics should be optimized with the initial additives such as CaO, MgO, and $Cr_3O_3$ additives. The present section reports the influence of CaO additives on the micro-structure, and optical properties of $Cr^{4+}$:YAG ceramics.

*2.2.1 The influence of CaO additive on sintering of YAG ceramics*

The first the influence of CaO sintering additives on the sintering trajectory should be considered. It should be noted that the presence of $Cr_2O_3$ additive in green body changes the sintering trajectory [19], and therefore the influence of the CaO on ceramics with and without chromium will be different. However, the general role of CaO on the ceramics densification can be extracted.

It was reported that the presence of CaO promotes densification of YAG ceramics, especially in a low temperature region [41] (Fig. 2.3). The positive influence of CaO additive on YAG ceramics densification might be explained thorough formation of liquid phases during the interaction of $CaO-Al_2O_3-Y_2O_3$. In fact, the interaction of CaO with $Al_2O_3$ can lead to formation of liquid phase at 1370 °C at right molar ratios [54]. The similarly, the liquid phase can be observed at 1400 °C in phase relations in the system $CaO-Al_2O_3-Y_2O_3$ [55]. Despite the interaction of $Al_2O_3$ and $Y_2O_3$ at the heating stage with the formation of YAP, YAM, and YAG phases, a fraction of $Al_2O_3$ and $Y_2O_3$ is still present at temperature 1400 °C [19] and can interact with CaO [40]. Several papers provide the experimental evidence of CaO interactions with $Al_2O_3$ during the sintering of YAG ceramics. The presence of calcium aluminate was detected in the Ca:YAG ceramics [36], but this result has been detected for ceramics after air annealing [36]. The air annealing caused migration of $Ca^{2+}$ ions from ceramics grain to aluminum inclusions [20]. The other paper showed evidence of liquid Ca-rich phase during the sintering of Cr:YAG ceramics but synthesized from YAG powders instead of $Y_2O_3$, $Al_2O_3$ oxides [38].

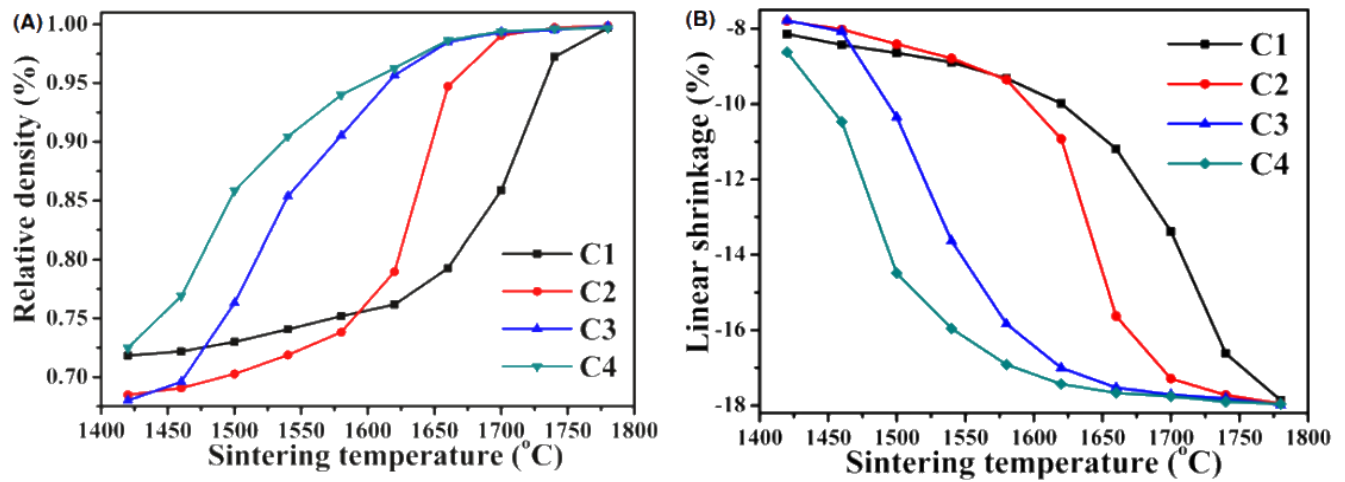

Fig. 2.3: Relative density (A) and linear shrinkage (B) vs temperature for Ca:YAG ceramics (C1 - 0.16 at.%, C1 - 0.32 at.%, C1 - 0.64 at.%, and C1 – 0.95 at.%) (Reproduced with permission from Ref. [41] © Wiley-Blackwell 2017).

Addition of CaO additives have influenced the sintering trajectory of YAG ceramics at the heating stage and followed isothermal annealing. This influence can be seen on the shrinkage curves measured for YAG and Ca:YAG ceramics ( Fig. 2.3) [36]. These curves can be separated on the two regions: first from ~1000 °C to ~1400 °C corresponds to the formation of YAP phase and second from ~1400 °C up to the isothermal annealing corresponds to formation of YAG phase (4.56 g/cm$^3$). Addition of Ca caused shifting of the maximum shrinkage value to lower temperature, which is found at 1300 °C, and decrease the shrinkage rate in temperature regions 1400 °C – 1650 °C ( Fig. 2.3) [36]. Moreover, the shrinkage rate became positive at temperature ~1450 °C for Ca-doped ceramics. It should be noted that the formation of most dense YAP phase (5.35 g/cm$^3$) (compared to initial $Y_2O_3$ (5.01 g/cm$^3$), $Al_2O_3$ (3.95 g/cm$^3$), and YAM (4.38

g/cm$^3$)) partially responsible for the maximum shrinkage at ~1300 °C. The positive shrinkage (expansion) at ~1450 °C is due to formation of more expanse YAG phase (4.56 g/cm$^3$) from YAP phase (5.35 g/cm$^3$). Therefore, the shifting in the maximum shrinkage and the appearance of positive shrinkage in Ca:YAG ceramics might be explained by increase in the rate of YAP→YAM→YAG phase transformations with the addition of CaO additive.

Incorporation of Ca$^{2+}$ ions in YAG lattice occurs instead of Y$^{3+}$ ions in dodecahedral site. Due to the difference in the state of valence, the charge compensation mechanism is required. incorporation of Ca$^{2+}$ ions occur as charge neutral [Ca$^{2+}$…$^1/_2$V$_O$] complex during vacuum sintering [19,36]. The incorporation of larger Ca$^{2+}$ (0.112 nm [56]) ion instead smaller Y$^{3+}$ (0.102 nm [56]) ion caused an increase in the lattice parameter which was detected earlier [36]. However, the lattice expansion is stopped if concentration of Ca$^{2+}$ ions exceed 0.05 at.%, therefore it was reported that 0.065(15) at.% of Ca$^{2+}$ ions could be proposed as the solubility limits of Ca$^{2+}$ ions in YAG lattice [36]. This value is in agree with the proposed earlier for YAG single crystal which is 0.08% at.% [57]. It is important that all calcium remains in the YAG lattice after vacuum sintering even for concentration of 0.5at.% [27]. Incorporation of large Ca$^{2+}$ ions instead small Y$^{3+}$ ions can cause decrease in the densification of YAG ceramics [36]. The densification of YAG ceramics during SSR sintering is controlled by the diffusion of the slowest cations – Y$^{3+}$ [58]. Formation of complex clusters with low mobility such as [Ca$^{2+}$…$^1/_2$V$_O$] deplete the concentration of cation, thus decrease the diffusion rate of slowest ions in YAG ceramics (Y$^{3+}$) resulting in reduction of the densification rate [36].

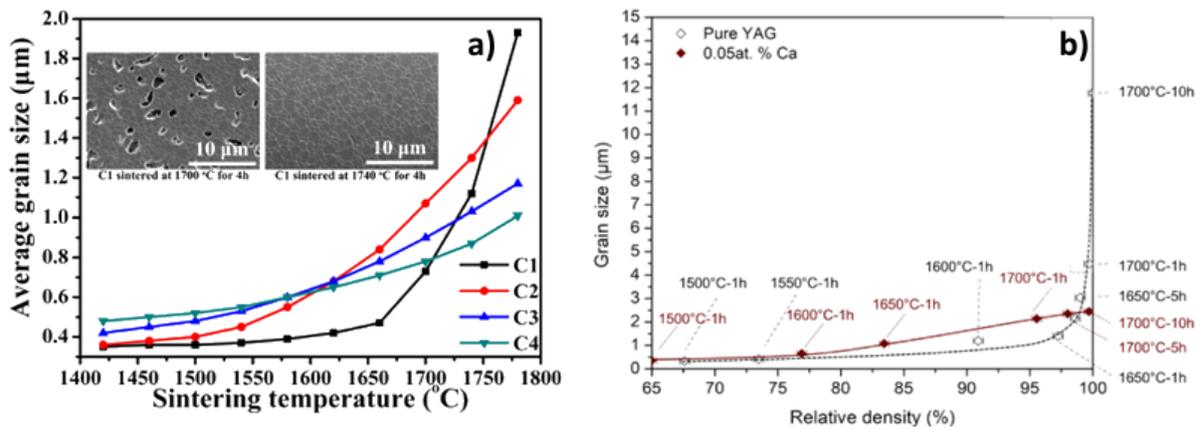

Fig. 2.4: a) Average grain size (C1 - 0.16 at.%, C1 - 0.32 at.%, C1 - 0.64 at.%, and C1 – 0.95 at.%) (Reproduced with permission from Ref. [41] © Wiley-Blackwell 2017), and b) sintering trajectories of Ca:YAG ceramics (Reproduced with permission from Ref. [36] © Elsevier BV 2023).

The main influence of the CaO additives on the ceramic sintering is the inhibition of grain growth [40,41]. The grain grow control occurs due to the segregation of $Ca^{2+}$ ions on the grain boundary [19,25,36], more detail sees in subsection 2.5.1. The increase in the concentration of CaO caused an increase in the average grain size up to the temperature of 1660 °C [41]. After 1660 °C, ceramics with the lower concentration of CaO exhibits faster growth (Fig. 2.4(a)) [41]. It should be noted that the grain size evolution depends on the sintering route and microstructure evolution of the ceramics. Therefore Ca:YAG ceramics sintered at temperature 1750 °C exhibits decrease [41] or increase [40] of grain size with increase in concentration of CaO depending on the sintering route. The average grain size of Ca:YAG ceramics [35,40,41] is less than in YAG [42] or Si:YAG [39] ceramics. Moreover, the often case when the abnormal grain grows was detected for undoped YAG ceramics [35,42]. YAG and Ca(0.05at.%):YAG ceramics synthesized in the same condition in vacuum at 1700 °C for 10 h have the average grain size ~12 µm and ~2 µm respectively (Fig. 2.4(b)) [36], proven the inhibition of grain growth in YAG ceramics by CaO additives. It is expected that the using of different source of

$Ca^{2+}$ ions such as CaO [24,34,35] or $CaCO_3$ [25,36] will have influence on the sintering trajectory of YAG ceramics. Unfortunately, it is difficult to extract the influence of chancing calcium source to $CaCO_3$ from the existing data.

*2.2.2 The influence of MgO additive on sintering of YAG ceramics*

In contrast to the CaO, MgO additives often used as sintering aids for YAG ceramics [35,42]. Using MgO as sintering additives was firstly reported in 1984, where translucent YAG ceramics was fabricated [59]. The influence of the MgO additive on the sintering of YAG ceramics can be described by the similar processes as that proposed for CaO additives. MgO additives can interact with the $Al_2O_3$ or/and $Y_2O_3$ creates an intermediate phase which can influenced the sintering trajectory of YAG ceramics. The interaction of CaO with $Al_2O_3$ or/and $Y_2O$ might cause formation of eutectic at temperature lower than ~1400 °C [54,55]. The interaction of MgO with $Al_2O_3$ or/and $Y_2O_3$ can form only a set of solid solutions without liquid phase at temperature lower than the sintering temperature [40,60].

Like the CaO additives, using MgO inhibits grain grow of the YAG ceramics [36]. The inhibition occurs due to the segregation of $Mg^{2+}$ on the grain boundary [25,36], details see in subsection 2.5.2. In some cases, high concentration of MgO additive can cause appearance of abnormal grain growth [61], but in most cases this was not the case [42]. The change of the concentration of MgO can slightly change the average grain size, which remains smaller than that for undoped YAG ceramics [35]. The average grain size of Mg:YAG ceramics was <5 μm while the average grain size of YAG ceramic sintered in the same condition was bigger than 10 μm [35]. However, the reported grain size for Mg:YAG ceramics differ significantly from ~1 μm [35] to 20 μm [42]. The inhibition of grain growth by $Mg^{2+}$ ions as well as $Ca^{2+}$ ions occur due to the segregation on grain boundary [20,36].

The change in the concentration of MgO additive has influenced the sintering trajectory of Mg:YAG ceramics. YAG ceramics with the higher concentration of MgO additive reaches the same density at lower temperature [61]. This influence might be explained by formation of the intermediate Mg-rich phases on the grain boundary [40,61], or/and formation of structural defects in YAG lattice [62]. Incorporation of $Mg^{2+}$ ions in YAG lattice occurs with the formation of $[Mg^{2+}\ldots{}^1/_2V_O]$ charge neutral complex [35] or/and by the self-compensation mechanism, occupying any cation site and an empty interstitial position [11]. The most of authors claim that the $Mg^{2+}$ ions substitute $Al^{3+}$ ions in octahedral site due to their similarities in the ionic radii [35,36,42]. However, effective ionic radii for $Y^{3+}$ ion (0.102 nm [56]) and $Mg^{2+}$ ion (0.089 nm [56]) in dodecahedral site differs on 13% while effective ionic radii for $Al^{3+}$ ion (0.054 nm [56]) and $Mg^{2+}$ ion $^+$ (0.072 nm [56]) in octahedral site differ at 25% [63]. It was reported that the $Mg^{2+}$ ions can replace $Al^{3+}$ ion as well as $Y^{3+}$ ion [11]. For some cases, the increase in the concentration of $Mg^{2+}$ ions caused expansion of YAG lattice indicating that larger $Mg^{2+}$ ion (0.069 nm [56]) replace smaller $Al^{3+}$ ion (0.054 nm [56]) [42]. On the other hand, there is paper reported opposite [63]. Depending on the concentration of MgO and the stoichiometry, $Mg^{2+}$ ions can replace $Y^{3+}$ ions or/and $Al^{3+}$ ions [63]. It should be noted that the deviation from YAG stoichiometry can be at least a half or two percent for excess of $Y^{3+}$ ions or $Al^{3+}$ ions without appearance of secondary phase [64].

The preferential positions for $Mg^{2+}$ ions and their solubility remain under question. Addition of more than 0.06 wt.% MgO (0.035at.% of $Mg^{2+}$ ions to $Al^{3+}$ ions) caused a drop in the lattice parameters and appearance of Mg-rich secondary phases indicating the solubility limit of $Mg^{2+}$ ions [42]. Cr,Ca,Mg:YAG ceramic with the Mg concentration 0.04 at.% shown segregation of magnesium in Mg-rich layer [25], indicating overwhelming the solubility limits. However, the part of $Mg^{2+}$ ions can be loosed due to evaporation [65], thus the reducing the value of determined earlier solubility limit of $Mg^{2+}$ (0.035at.% [42]). For example, the sintering of YAG ceramics in wet hydrogen at 1800 °C caused lose 80% of $Mg^{2+}$ ions while the concentration of

CaO does not change markedly [65]. The other work reports that the concentration of $Mg^{2+}$ ions do not excess 0.005 at.% for replacing $Y^{3+}$ ions or/and $Al^{3+}$ ions [63].

*2.2.3 The mutual influence of MgO and CaO additives on optical properties of YAG ceramics*

Despite the similarities, the main difference between CaO and MgO additives is the negative effect of $Ca^{2+}$ ions on densification kinetics of YAG ceramics [36] while the opposite influence has $Mg^{2+}$ ions [42]. Moreover, densification rate of Mg(0.5at.%):YAG ceramics is higher than for Ca(0.5at.%):YAG ceramics [40], consequently MgO can be used as efficient sintering aid for YAG ceramics [36,42,49]. Therefore, the frequent practice is the use of both $Mg^{2+}$ and $Ca^{2+}$ ions for synthesized of $Cr^{4+}$:YAG ceramics. In this subsection we discussed the influence of CaO and MgO additives on the sintering processes of YAG ceramics.

$Me^{2+}$:YAG ceramics synthesized in the same conditions exhibits influence of the Ca/Mg ratio on the microstructure. As can be seen on Fig. 2.5, Me:YAG ceramics (1780 °C at 4h) with the Ca/Mg ratio from Ca-doped only to $^4/_1$, $^3/_2$, $^2/_3$, $^1/_1$, and finally Mg doped only had the average grain size – 1,7 μm, 1,9 μm, 2,2 μm, 2,8 μm, 3,5 μm, and 4,5 μm, respectively [35]. It is indicating that CaO additives more efficient inhibiting grain growth than MgO additive. It should be noted that the MgO additive is still efficient in grain growth. Ceramics sintered without any sintering additives had larger grains and exhibits abnormal grain grow regime when temperature exceed 1650 °C [35].

The change in the Ca/Mg ratio also has influence on the sintering trajectory of Ca,Mg:YAG ceramics. Below the density of 98%, the relatively density of the Ca,Mg:YAG ceramics increase with the increasing amount of MgO additive ( Fig. 2.5(b)) [35]. A similar pattern was detected for Mg:YAG ceramics [61]. The undoped ceramics have higher density (below 98%) at each sintering temperature than in compare with the doped samples. It should be noted that commonly most of the YAG ceramics reach the theoretical density above 98% at a sintering temperature 1750 °C or higher, therefore it is not so critical how fast the densification occurs

[35,36]. Even small concentration of pores will significantly decrease the transmittance of the ceramics. For example, 10 mm thick sample with the residual porosity 0.002 vol.% and 0.02 vol.% have the transmittance at 630 nm ~ 60% and ~10%, respectively.

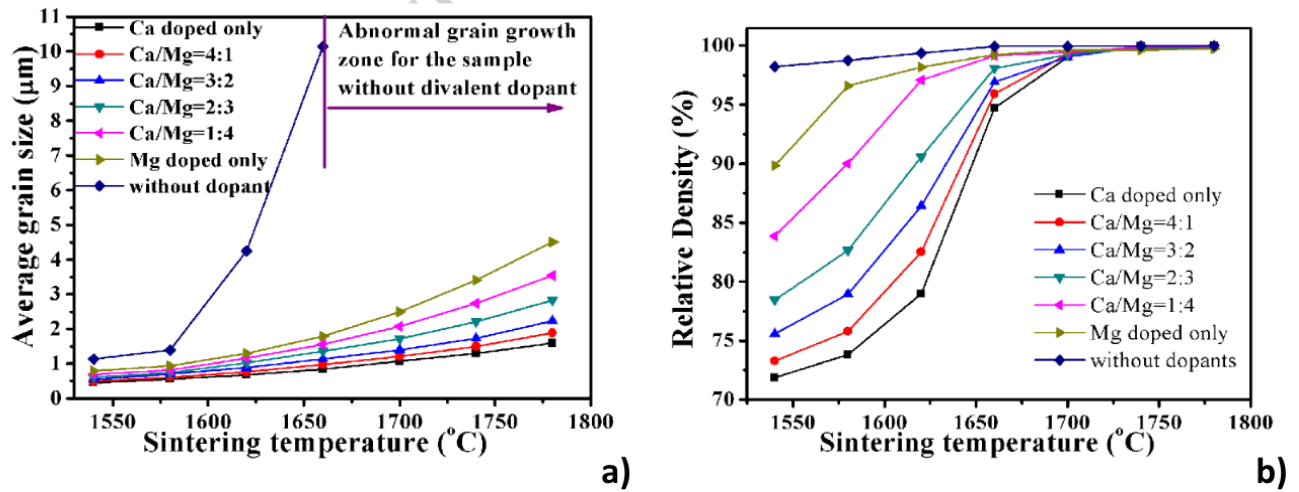

Fig. 2.5: a) Average grain size and b) relatively density of Ca,Mg:YAG ceramics (Reproduced with permission from Ref. [16] © Elsevier BV 2017).

The change in Ca/Mg ratio changes the pore evolution of Ca,Mg:YAG ceramics. YAG ceramics with only MgO additives exhibit more rapid pore evolutions that sample with CaO additives [35] indicating difference in the microstructure evolutions ( Fig. 2.5(b)). During the densification, microstructure went through – individual particles → network of open pores → closed pores → full dense ceramics. Probably that the lower density of ceramics with higher CaO concentration is due to the later collapse of open pores. It is important to remain the network of open pores as small as possible because in this case the chains of small, closed pores will be formed. Smaller pores can be easily eliminated from the ceramic volume than larger one [19]. Earlier collapse of network of open pores as for Mg:YAG or later collapse as for Ca;YAG can cause appearance of large, closed pores. Tuning the Ca/Mg ratio allows us to reach optimal microstructure evolutions resulting in formation of high transparent ceramics, as for example

$1/4$ ratio of Ca/Mg for Ca,Mg:YAG [19]. It should be noted that this ratio can be different for ceramics prepared and sintered in different conditions than for [19].

**2.3 The influence of CaO and/or MgO additives on the sintering of $Cr^{4+}$:YAG ceramics**

The possibility of synthesized transparent YAG ceramics using CaO or/and MgO additives [41,61] opens road towards the sintering of $Cr^{4+}$:YAG ceramics. Fig. 2.6 shown an example of mirror polished $Cr^{4+}$:YAG ceramics. The section above describes the way to sintering high transparent Me,Cr:YAG ceramics (Me – $Ca^{2+}$ or/and $Mg^{2+}$). It should be noted that the addition of chromium additives changes the sintering trajectory of ceramics [19], and therefore the sintering parameters for Cr,Me:YAG can be different as for Me:YAG. This section collects the information about the mutual influence of the MeO and $Cr_2O_3$ additives on sintering trajectory of $Cr^{4+}$:YAG ceramics.

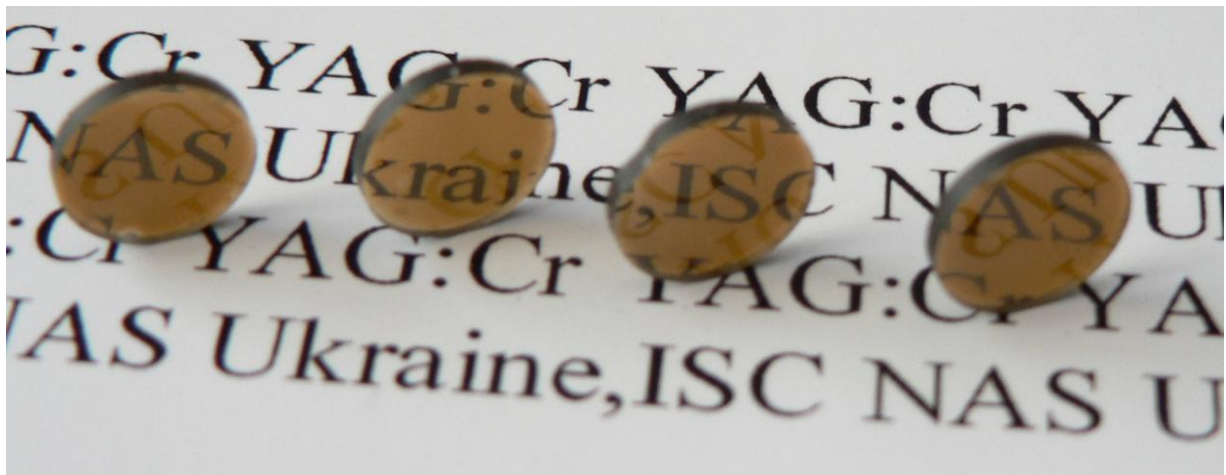

Fig. 2.6: Photograph of the mirror polished Cr,Ca:YAG ceramics.

*2.3.1 The influence of CaO additive on sintering of $Cr^{4+}$:YAG ceramics*

Due to the success in sintering Cr,Ca,Mg:YAG transparent ceramics [23], there is limited data related to the influence of CaO additives in the formation of Cr:YAG ceramics. Moreover, MgO is preferential for synthesis of transparent garnet ceramics [35]. This do not mean that CaO is not suitable as single additive for sintering $Cr^{4+}$:YAG ceramics, the opposite is true [10,22,27].

The key to sintering high quality Cr,Ca:YAG ceramics is to turn the concentration of both CaO, $Cr_2O_3$ additives and their ratio. It was shown that the change in the ratio of Cr/Ca from $^4/_1$, $^1/_1$, $^1/_3$ to $^1/_5$ caused change in the transparency and microstructure of the Cr,Ca:YAG ceramics [27]. The microstructure of the synthesized ceramics depends on the Cr/Ca ratio. Cr(0.1at.%),Ca(0.8at.%):YAG ceramics ($^1/_5$ Cr/Ca ratio) characterized by the presence of Ca-rich thin film on the surface of ceramics after air annealing ( Fig. 2.8(b)) while no such film was found for the others Cr/Ca ratio. The same case was found for the ceramics with TEOS sintering additives (subsection 2.1.1, Fig. 2.1(c)) [34] which indicates the doping over the solubility limit for $Ca^{2+}$ ions ( Fig. 2.8(b)). More details see in section 2.4. The change in Ca concentration have influence on the average grain size and porosity of the ceramics. For example, Cr,Ca:YAG ceramics synthetized at 1750 °C with Cr/Ca $^4/_1$, $^1/_1$, $^1/_3$ and $^1/_5$ had average grain size 1 μm, 1.3 μm, 1 μm and 0.8 μm respectively [27]. The presence of CaO additives inhibits grain growth [40,41], compared to undoped or Mg-doped YAG ceramics. However, the increase of the concentration of CaO might decrease [41] or increase [40] the average grain size. In the case of Cr,Ca:YAG ceramics the average grain size nonlinearly changes with CaO amount [27]. Moreover, the average grain size of Cr(0.1at.%),Ca(0.5at.%):YAG ceramics ($^1/_3$ Cr/Ca ratio) was larger than for Ca(0.5at.%):YAG synthesized in the similar conditions [40] ~1 μm and ~3.5 μm, respectively. This indicates that the addition of $Cr_2O_3$ additives change the sintering trajectory of the ceramics [19].

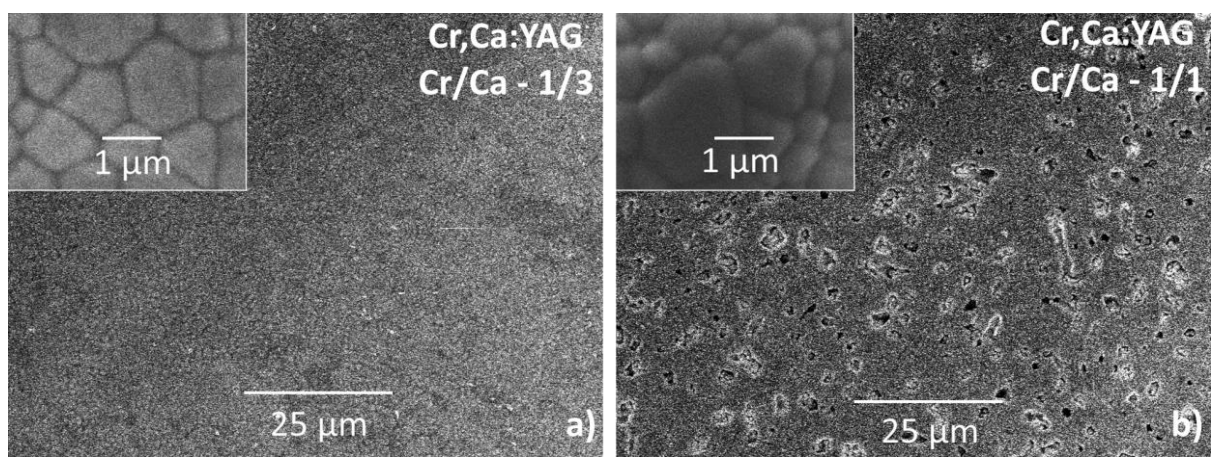

Fig. 2.7: SEM image of Cr,Ca:YAG ceramics synthesized at 1750 °C for Cr/Ca ratio a) $^1/_3$, and b) $^1/_1$.

Special attention should be paid to the Cr,Ca:YAG ceramic with the $^1/_1$ Cr/Ca ratio. These ceramics have higher average grain size and characterized by presence of large pores in the size of few micrometers (Fig. 2.7). Moreover, the same composition synthesized in the higher temperature (1800 °C) characterized by the presence of abnormal grain growth, which was not found for the other compositions (Fig. 2.8). The abnormal grains characterized by high intergrain porosity with the pore diameters up to ten micrometers. This indicates that the abnormal grain growth occurs at the heating stage with the fast collapse of network of open pores. In the same ceramics there are the areas with normal growth characterized by normal grain distributions with the high density and in-line transmittance ~ 80% at 1100 nm (see insert on Fig. 2.8(a)). Abnormal grain growth is usually present in the samples without any additives [35,42] or in Cr,Ca:YAG ceramics where all CaO was bounded with $SiO_2$ forming intermediate phases [34].

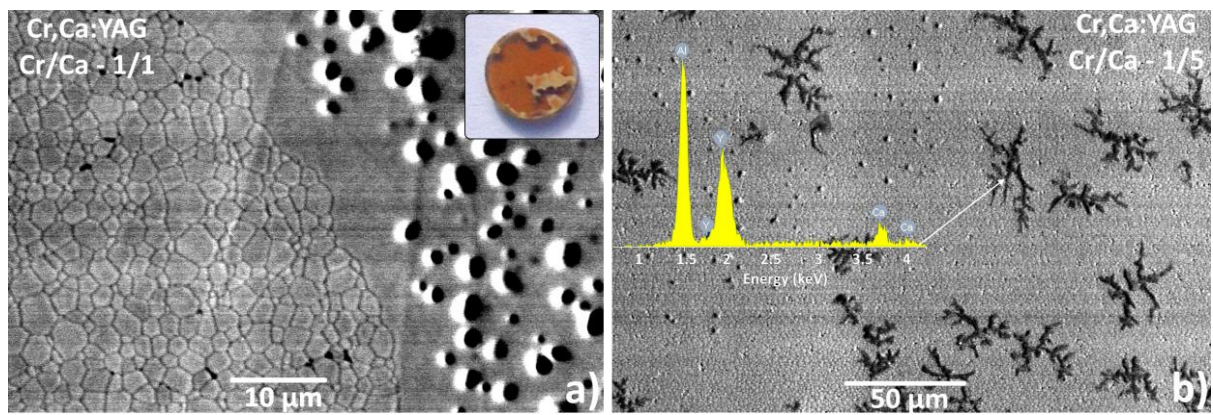

Fig. 2.8: SEM image of Cr,Ca:YAG ceramics synthesized at 1800 °C for Cr/Ca ratio a) $^1/_1$, and b) $^1/_5$.

The possible explanation of the difference in the properties of Cr,Ca:YAG ceramics is interaction between CaO and $Cr_2O_3$. The microstructure of the Cr,Ca:YAG ceramics correlates with the formation of possible intermediate phases due to the interaction of CaO and $Cr_2O_3$ according to their phase diagram [66]. $Cr_2O_3$ is present in the ceramics at least at 1100 °C [19], therefore it is possible the interaction between CaO and $Cr_2O_3$ at lower temperature. The liquid phase can be found 1022 °C for Cr/Ca ratio $^1/_1$, while the using higher ratio shifts the temperature of formation of liquid phase to 1275 °C. The absence liquid phase arose for Cr/Ca ratio $^4/_1$. Based on this we can conclude that the interaction between CaO and $Cr_2O_3$ for Cr/Ca ratio $^1/_1$ caused occurrence of liquid phase at temperature ~ 1020 °C. This changes the sintering trajectory of the Cr,Ca:YAG ceramics. The densification begins at the temperature of 1000 °C and higher, therefore formation of liquid phases might accelerate the densification process. Faster densification might cause the earlier collapse of open pores into the large, closed pores with the diameters more than several microns ( Fig. 2.8(a)). After exceeding the temperature 1100 °C, most of $Cr_2O_3$ dissolves in $Al_2O_3$ or forms the other phases with $Y_2O_3$ [19]. At the temperature of formation of liquid phase in CaO-$Cr_2O_3$ system for Cr/Ca ratio $^1/_3$, or $^1/_5$ (1275 °C) most of $Cr_2O_3$ is unavailable. Therefore, the ceramics with this Cr/Ca ratio have the same microstructure. The exception is the presence of Ca-rich thin film on the ceramic surface ( Fig. 2.8(b)).

Sines the $Cr_2O_3$ is disappears at the temperature of 1100 °C by dissolution in $Al_2O_3$ and followed interaction with $Y_2O_3$, it is unclear why the average grain size of Ca:YAG ceramics [40] differs from the grain size of Cr,Ca:YAG ceramics [27] sintered in the same condition. The possible explanation might be the interactions of chromium oxides with the other oxides. The interaction of $Cr_2O_3$ with $Y_2O_3$ results in the formation of thin film of $YCrO_3$ on the grain boundary [19]. Formation of this these films can change the interaction of CaO with the YAG grains resulting in collections of more CaO at grain boundary to higher temperatures than compared to Ca:YAG ceramics. It should be noted that the proposed explanation is based on the assumptions and simplifications such as using $CaO-Cr_2O_3$ phase diagram. Then real behavior might be more complex, but this shows the general trend in Cr,Ca:YAG ceramic sintering. More details see in the section 2.4.

*2.3.2 The influence of MgO additive on sintering of $Cr^{4+}$:YAG ceramics*

After TEOS, MgO is the second most used sintering aid for creates luminescence-doped YAG transparent ceramics [4,6,30,49,62]. However, it is a limited data regarded to the sintering of Cr,Mg:YAG ceramics, with few papers report so far [52,67]. The presence of $Cr_2O_3$ additive changes the sintering trajectory of YAG ceramics [19], so it is hard to use the data collected for Mg:YAG ceramics. Moreover, the presence of $Cr^{3+}$ ions increase solubility of $Mg^{2+}$ ions in YAG lattice [68] which is indicates the difference in the sintering of Mg:YAG and Cr,Mg:YAG ceramics. For ceramics sintering in similar conditions, in was reported presence of Mg-rich phases for the Mg(0.5at.%):YAG ceramics [40], while the Cr(1at.%),Mg(0.5at.%):YAG ceramics have absence of secondary phases [52].

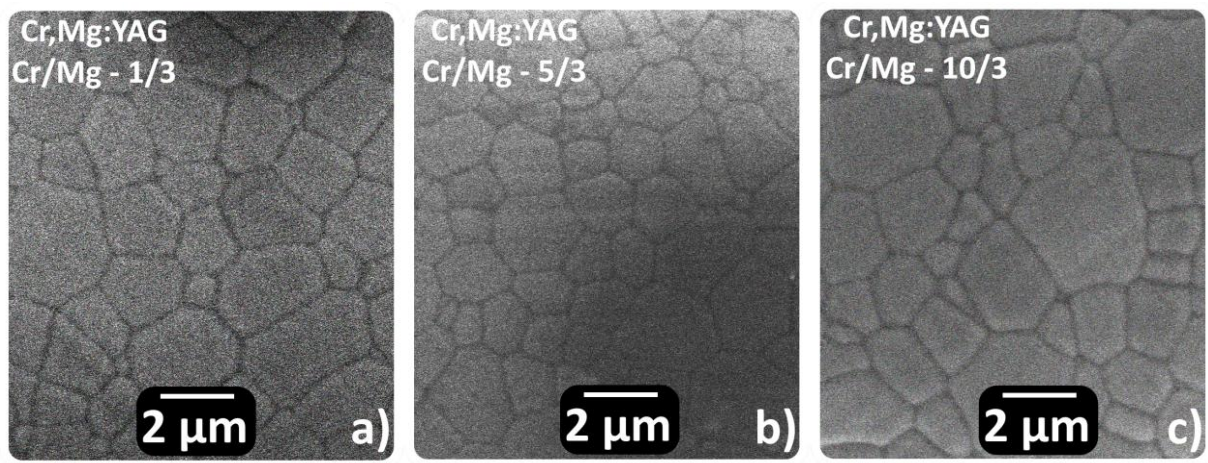

Fig. 2.9: SEM image of Cr,Mg:YAG ceramics synthesized at 1765 °C for Cr/Mg ratio a) $^1/_3$, and b) $^5/_3$, and c) $^{10}/_3$, respectively.

In contrast to the Cr,Ca:YAG ceramics, the change of Mg/Cr ratio from $^1/_3$ to $^5/_3$, and $^{10}/_3$ have little influence on the microstructure and transparency ( Fig. 2.10) [67]. For the Cr,Ca:YAG ceramics synthetized at 1750 °C, the average grain size was 2.1(1) μm, 1.8(1) μm, and 2.2(1) μm, for Mg/Cr ratio $^1/_3$, $^5/_3$, and $^{10}/_3$, respectively ( Fig. 2.9). The Mg(0.5at.%):YAG ceramics with the same concentration of $Mg^{2+}$ ions and sintered in the same condition have the average grain size 3.2(2) μm [40], which is half larger than for Cr,Mg(0.5at.%):YAG. This indicates that mutual addition of MgO and $Cr_2O_3$ additives more efficiently inhibit the grain growth. The inhibition of grain growth occurs due to accumulation of MgO on the grain boundary, which decreases their mobility [36]. The presence of $Cr^{3+}$ ions in YAG lattice changes the distribution of MgO [68], which might be an explanation of reduction in grain size. A major fraction of $Mg^{2+}$ ions can be lost due to the evaporation [65]. The decrease in average grain size might be explained by decrease in evaporation rate of MgO. The presence of $Cr^{3+}$ ions in YAG lattice changes the distribution of $Mg^{2+}$ ions [68], thus can reduce the evaporation rate of the last. The influence of the $Cr_2O_3$ additives is stronger for garnet ceramics doped with the MgO additive above the solubility limits. The ceramics synthesized with the small concentration of MgO exhibits the little change in the average grain size with the addition of $Cr_2O_3$ (Cr(0.1at.%),Mg(0.05at.%):YAG - 3.9(3) μm [26], Mg(0.05at.%):YAG - 3.2(2) μm) [40]).

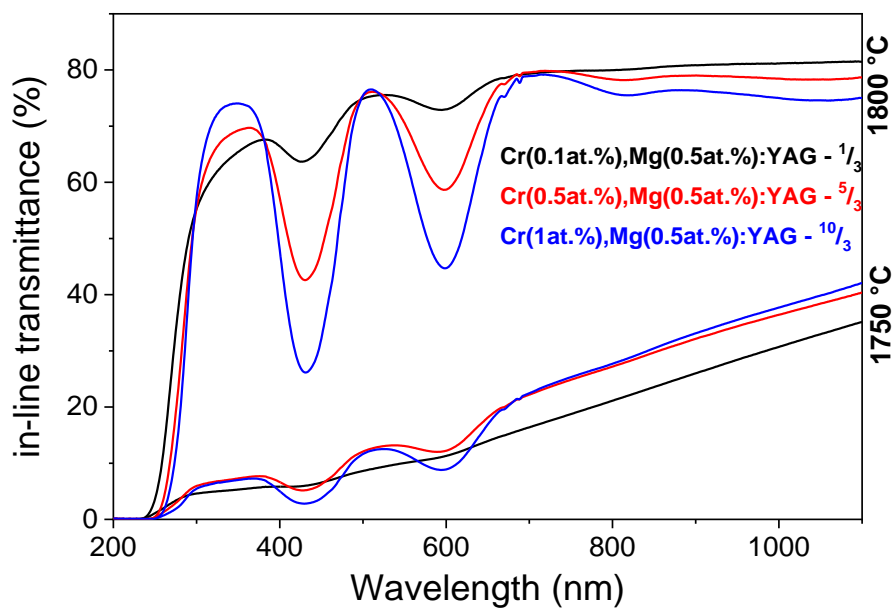

Fig. 2.10: In-line transmittance spectra of Cr,Mg:YAG ceramics with Cr/Mg ratio $^1/_3$, $^5/_3$, and $^{10}/_3$, synthetized at 1750 °C and 1800 °C, data was modified from [67].

Apart from absorption from chromium ions, in-line transmittance of the Cr,Mg(0.5at.%):YAG ceramics is close for different Mg/Cr ratio. This indicates that interactions between MgO and $Cr_2O_3$ additives with the formation of low temperature liquid phases do not occur or insignificant. This correlates with the MgO and $Cr_2O_3$ phase diagram [69] which showed the absence of liquid phase in temperature region lower than 2000 °C. However, the microstructure of the Cr,Mg(0.5at.%):YAG ceramics depends on the temperature of the vacuum sintering ( Fig. 2.10). The change of the temperature of the vacuum sintering at 15 °C from 1750 °C to 1765 °C caused decrease residual porosity from 0.015 vol.% to ~ 0.001 vol.% caused rise of in-line transmittance from ~40% to ~80% at 1100 nm [52]. This is agreed with the data reported for Mg,Yb:YAG ceramics, where change on 10 °C caused drop in transparency at ~25% [70].

*2.3.3 The mutual influence of MgO and CaO additives on sintering of $Cr^{4+}$:YAG ceramics*

The most progress related to the sintering $Cr^{4+}$:YAG ceramics was with the use both CaO and MgO additives [17,23]. In fact, the first paper on the $Cr^{4+}$:YAG ceramics was done with both CaO and MgO additives [14]. The presence of MgO sintering additives allows to synthesize transparent ceramics while the CaO additive responsible for recharge chromium into tetravalent state. Of course, transparent $Cr^{4+}$:YAG ceramics can be obtained with only CaO additive [27], or high concentration of $Cr^{4+}$ ions can be obtained with only MgO additive [52].

With some exceptions [22], the most of papers used Ca/Mg ratio $^1/_1$ [16,17,23], therefore the followed discussion based on the results obtained for this ratio. By varying Me concentrations, the change in the Cr/Me (Me – Ca+Mg) ratio changes the microstructure of the obtained $Cr^{4+}$:YAG ceramics. The decrease in Cr/Me ratio from $^1/_2$ to $^1/_4$ caused an increase in average grain size in half from 3.1 μm to 4.7 μm, respectively [16] ( Fig. 2.11(a,b)). The further decrease in the Cr/Me ratio to $^1/_6$ caused abnormal grain growth with the capturing of pores inside the grain volumes [16] ( Fig. 2.11(c)). In contrast to the Cr,Mg:YAG ceramics [52], the same behavior is typical for Cr,Ca:YAG ceramics, where the change in Cr/Ca ratio caused change in microstructure [27]. This indicates that the change in microstructure with the change in Cr/Me ratio is due to the influence of CaO additive. However, the abnormal grain growth for Cr(0.1at.%),Ca,Mg:YAG ceramics was found at Cr/Me ratio at $^1/_6$ (Cr/Ca ratio at $^1/_3$) [16], while for Cr(0.1at.%),Ca:YAG ceramics abnormal grain growth was found for Cr/Ca ratio $^1/_1$ [27] ( Fig. 2.8(a)). Such difference can be explained by the influence of the MgO additive, the presence of which changes sintering trajectory of YAG ceramics [35].

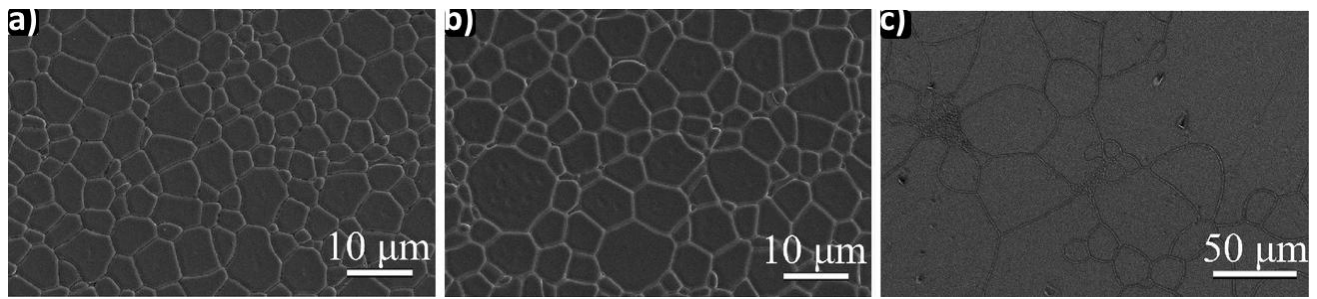

Fig. 2.11: SEM image of Cr,Ca,Mg:YAG ceramics synthesized at 1800 °C for Cr/Me ratio a) $^1/_2$, and b) $^1/_4$, and c) $^1/_6$, respectively (Reproduced with permission from Ref. [16] © Wiley-Blackwell 2015).

The different behavior was found for the Cr,Ca,Mg:YAG ceramics with the same concentration of CaO and MgO (0.2 mol.% to $Al^{3+}$ ions each) and different concentration of $Cr^{3+}$ ions (0.05-0.3 at.%) [23]. Except for $Cr^{3+}$ absorption bands, the transparency remains the same [23] (Fig. Fig. 2.12), indicating the same microstructure evolutions of this ceramics. The change in Cr/Me ratio from $^{0.5}/_4$, $^1/_4$, $^2/_4$, to $^3/_4$ caused an increase in the average grain size from 2.7 μm to 1.8 μm [23]. Increase of concentration of $Cr^{3+}$ ions in YAG lattice caused redistribution of both $Ca^{2+}$ and $Mg^{2+}$ ions [68], which is the reason for change in the average grain size. Most probably that the microstructure of Cr,Ca,Mg:YAG ceramics determinate by the $Mg^{2+}$ ions. This conclusion was made from the reported microstructure of Cr,Ca,Mg:YAG, Cr,Mg:YAG, and Cr,Ca:YAG ceramics synthesized in the same conditions where the average grain size was 3.9(3) μm, 3.9(3) μm, and 1.9(2) μm, respectively [26]. The presence of MgO additives caused the growth of grain sizes twice [26], however, the other work does not confirm it and reported the same average grain sizes for both Cr,Ca,Mg:YAG, and Cr,Ca:YAG ceramics [22].

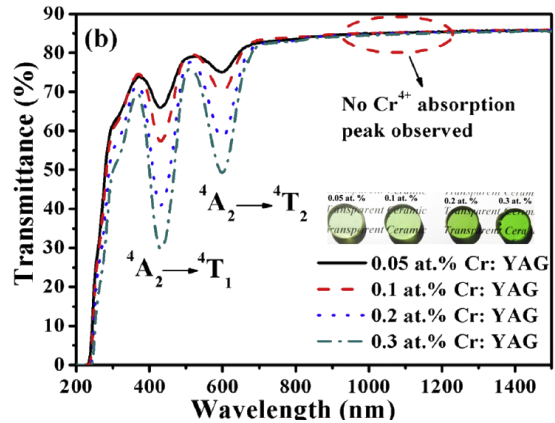

Fig. 2.12: optical in-line transmission spectra (b) of mirror polished Cr,Ca,Mg:YAG ceramics Cr/Me ratio - $^{0.5}/_4$, $^1/_4$, $^2/_4$, $^3/_4$ (Reproduced with permission from Ref. [23] © C Elsevier BV 2017).

**2.4 The influence of $Cr_2O_3$ additive on the sintering of $Cr^{4+}$:YAG ceramics.**

As was highlighted in the previous sections, the presence of $Cr_2O_3$ additive changes the sintering route of $Cr^{4+}$:YAG ceramics. As for example, Fig. 2.13 shown photo of Cr(0.1at.%),Ca(0.5at.%):YAG ceramics and Ca(0.5at.%):YAG ceramics sintered in the same conditions. The Ca(0.5at.%):YAG ceramics demonstrates worse transparency compared with Cr-doped ceramics ( Fig. 2.13). The in-line transmission of Cr,Ca:YAG and Ca:YAG ceramics is 82.4 and 65 % at 1100 nm respectively [19]. This indicates that $Cr_2O_3$ additives have an influence on the sintering process of YAG ceramics.

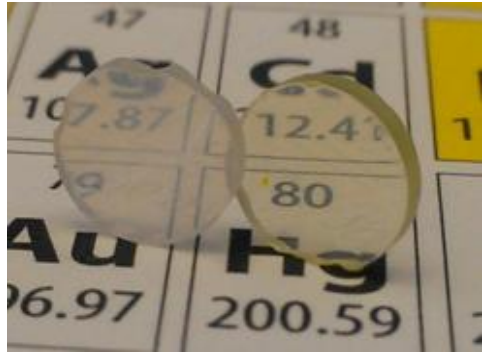

Fig. 2.13: Photo of the Ca(0.5at.%):YAG (left) and Cr(0.1at.%),Ca(0.5at.%):YAG (right) ceramics [19].

At the earlier stage of the ceramics sintering even before 1100 °C [19], a major fraction of $Cr_2O_3$ already dissolved in $Al_2O_3$. Moreover, the majority of $Cr^{4+}$:YAG ceramics were synthetized with the small concentration of $Cr_2O_3$ additive (0.1 at% to $Al_2O_3$) [17,23,26,27]. Therefore, the influence of the $Cr_2O_3$ additive on the sintering process is quite surprising. Most of the explanations regarded to the influence of the additives on the sintering processes are based on the formation of defects in the YAG lattice which improves the densification process [32,41], or presence of intermediate liquid phases during the interactions of functional additives on the grain boundary [27,34]. Since the chromium and aluminum ions have the same valence state, incorporation of $Cr^{3+}$ ions in YAG lattice occurs without formations of any defects in the YAG structure [10,20,71]. Therefore, it was proposed that the presence of the $Cr_2O_3$ additives caused appearance of the $YCrO_3$ phase which change the sintering trajectory of the YAG ceramics.

*2.4.1 The influence of the phase formations on the densification process.*

For understand the influence of the $YCrO_3$ phase on the sintering of YAG ceramics, there important to describe the processes occurs during the heating stage. In simplified cases, vacuum sintering can be separated on the two stages: heating and isothermal annealing. During the heating stage the density of YAG ceramics increase from of that of green body (50-70%) above

99%, while remaining present eliminated during the isothermal annealing [41,61]. Therefore, the key to the good optical quality YAG ceramics lies thorough engineering of microstructure evolutions of YAG ceramics to achieve the porosity and pores diameters as small as possible before the isothermal annealing. Pores in YAG ceramics can be considered as impurity phases of void. These voids can be removed from the ceramics thorough "evaporations" of "atoms of voids" (vacancies) and their diffusion to the absorber of atoms of void (grain boundary, edge of ceramics, etc.). The pores with the smaller diameters have a higher rate of evaporation than in compared to the larger one at the same conditions. Also, the closer pores to the absorber, the higher rate of evaporations. Therefore, it is important to keep the pores as small as possible and avoid of formation of intragranular pores.

The evolution of microstructure of ceramics from the green body to full dense ceramics goes through several steps. The classical densification curve remains s-shape where the highest densification occurs during the fast shrinkage stage. During the fast shrinkage stage densification occurs initially due to mutually moving and rotation of the powder particles itself caused by temperature gradients due to the heating of the ceramics. After some time, the moving and rotation of the powder particles slows down due to formation of connections and 3D network between the particles of the original powders. Driven by the mass transfer between the particles, the voids became take shape of long connected channel forming the network of open pores. This network of open pores collapses into the chains of the closed pores which then can be eliminated during the isothermal annealing. The wrong evolution of microstructure at any transition stage (individual particles → 3D network → open pores → closed pores) affects the final properties of ceramics. The earlier collapse of larger open pores caused formation of large, closed pores, as well as earlier connection of particles into 3D network led to the formation of wider network of open pores with high inhomogeneous in the diameters of the channels.

The main feature of the sintering of YAG ceramics by SSR sintering is that together with the densification, the solid-state reactions between initial compound occurs with the formations of intermediate phases with different density. This phase transformation is crucial to obtain transparent ceramics, proven of which is the fact that the sintering of YAG transparent ceramics from garnet powders is more challenge. It is not coincidence that the faster shrinkage rate (1100 °C – 1400 °C) occurs at the same temperature as the phase transformation (1000 °C – 1500 °C). Formation of more dense YAP phase (1100 °C – 1300 °C) allow to break connections between the particles which accelerate the densification thorough the moving and rotation of powder particles. The earlier or later formation of YAP phase will occur when this network is not present or evolve enough to be unbraked thus reducing the positive influence of the phase transformation on the densification.

*2.4.2 The influence of the chromium on the phase transformations during the sintering of YAG ceramics.*

The presence of $Cr_2O_3$ additives change phase transformations rates during SSR sintering, shifting formation of the YAP phase to the higher temperature. The interaction of the $Y_2O_3$-$Al_2O_3$ powders during SSR sintering resulted in the following phase transformations: $Y_2O_3$ + $Al_2O_3$ → $Y_4Al_2O_9$ → $YAlO_3$ → $Y_3Al_5O_{12}$ ( Fig. 2.14). Despite the small concentration the presence of $Cr_2O_3$ additive decreases the rate of the phase transformation, as was shown in the case of Cr(0.1at.%),Ca(0.5at.%):YAG and Ca(0.5at.%):YAG ceramics [19]. For example, during the heating stage at the temperature 1300 °C, ceramics with chromium have half lower concentration of YAP phase than compared to the ceramics without chromium [19].

Chromium is taken in according to replace aluminum and at the temperature of 1100 °C, most of the chromium is dissolved into $Al_2O_3$. According to the $Cr_2O_3$-$Y_2O_3$ phase diagram [72], the interaction between $Cr_2O_3$ and $Y_2O_3$ is different from between $Al_2O_3$ and $Y_2O_3$. Only one compound $YCrO_3$ is presented on the phase diagram of the $Cr_2O_3$-$Y_2O_3$ system with a negligible mutual solubility of $Cr_2O_3$ and $Y_2O_3$. Therefore, the interaction of $Cr:Al_2O_3$ with the $Y_2O_3$ leads to formation of YAM and $YCrO_3$ phases. $YCrO_3$ can precipitate as thing flakes along the grain boundaries of $Y_2O_3$ [73] during diffusion of $Cr^{3+}$ ions. Emerged $YCrO_3$ promotes inward oxygen diffusion and slows cation diffusion [73]. It should be noted that the presence of $YCrO_3$ was shown indirectly from the optical properties of Cr(0.1at.%),Ca(0.5at.%):YAG ceramics at the different temperatures of the heating stage [19].

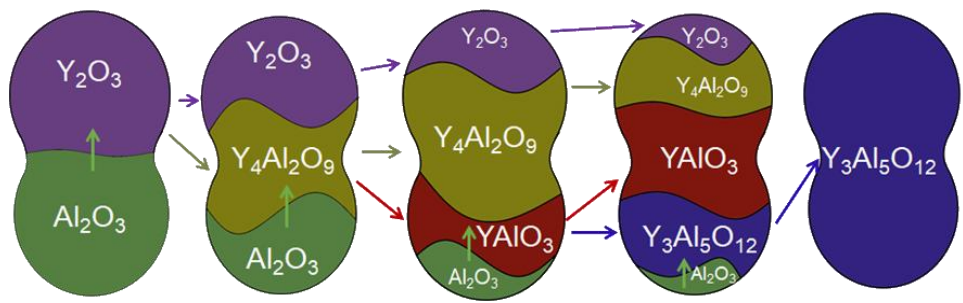

Fig. 2.14: The solid-state reactions of $Y_2O_3$ and $Al_2O_3$ particles.

Since the phase transformation plays a key role on the microstructure evolutions, the shift in the phase transformation for the chromium doped ceramics changes the spinning route of Cr(0.1at.%),Ca(0.5at.%):YAG ceramics [19]. The appearance of new intermedial phase causes the decreasing the contribution of SSRs in total shrinkage of the green body under the fast shrinkage stage that enhances the optical properties of the final ceramics [19]. For examples, the ratio between volume change due to SSR to volume change due to shrinkage for Cr(0.1at.%),Ca(0.5at.%):YAG and Ca(0.5at.%):YAG ceramics during the heating stage at 1300 °C is 0.21, and 0.54, respectively [19]. At the specific composition and sintering parameters, such changes promote the formation of the more homogeneous 3D network and results in Cr:YAG ceramic with enhanced optical properties.

The influence of the $Cr_2O_3$ additive in the sintering of Cr;YAG ceramics is due to the change in the relation between volume change due to SSR to volume change due to shrinkage resulting in change of sintering route. However, both SSR rates and shrinkage depends on the change in heating rate [74]. Also, analyzing the densification curves of Ca:YAG ceramics [36], it can be concluded that the different concentration of CaO additive have influence of both the SSR rates and shrinkage. This indicates that the influence of chromium can be different or even opposite for different Cr,Me:YAG ceramics with different composition or/and sintering parameters such as heating rate, isothermal annealing temperature, etc. The whole sintering process should be tunned separately for YAG ceramics doped with chromium.

**2.5 The role of $Me^{2+}$ ions (Me - Ca, Mg) in the formation of high optical quality $Cr^{4+}$:YAG ceramics**

The challenge in the manufacturing technologies of $Cr^{4+}$:YAG ceramic is to control the effect of divalent metal ($Me^{2+}$) additives required to compensate for the extra charge generated in the YAG lattice due to appearance of $Cr^{4+}$ ions. These additives can decrease transparency and change the ceramics optical absorption at 1.3-1.6 µm. The distribution and concentration of $Cr^{4+}$ ions are strongly influenced by the divalent additives, such as $Ca^{2+}$ or $Mg^{2+}$ ions [22]. The highest concentration of $Cr^{4+}$ was obtained under 3-4 times excess of calcium with respect to total chromium [27]. Such calcium excess raises two opposite problems that should be addressed by the technologists. An increase in calcium concentration is required to raise the concentration of $Cr^{4+}$ ions in ceramics, but a high calcium concentration might cause the precipitation of extra phases on the ceramic grain boundaries which deteriorate the transparency of the $Cr^{4+}$:YAG ceramics [34].

The optical properties of YAG-doped ceramics depend much on the final microstructure. Ions segregation at the grain boundaries affects the both for porosity elimination and grain growth [27,36]. The $Ca^{2+}$ migration forward to or backward from the ceramics grain boundaries can play a key role to obtain high optical quality $Cr^{4+}$,Me:YAG ceramics. Vacuum sintering and air annealing are the two basic steps in ceramics preparation by solid state reactive sintering, which determines the final calcium distribution in the ceramics. During the vacuum sintering $Me^{2+}$ ions can be stabilized by oxygen vacancies, while after air annealing part of $Me^{2+}$ ions recharge chromium in tetravalent state. The $Me^{2+}$ migration across the grains contributes to the concentration and distribution of $Cr^{4+}$ ions. In this section, the distribution of $Me^{2+}$ ions in $Cr^{4+}$,Me:YAG ceramics grain boundaries were discussed [20,36].

*2.5.1 Distribution of $Ca^{2+}$ ions in $Cr^{4+}$:YAG ceramics.*

An inhomogeneous distribution of $Me^{2+}$ ions was revealed in the Cr,Me:YAG [20,25] and Me:YAG [36] ceramics. By monitoring the lattice parameter of Ca:YAG ceramics, it was concluded that the solubility of Ca in YAG ceramics is 0.065(15) at.% [36] which corresponds to the reported value for YAG single crystal (0.08% at.%) [57]. The $Ca^{2+}$ ions distribution in the Ca(0.05at.%):YAG ceramics doped below solubility limit is more or less homogeneously. However, the calcium prefers to be near the grain boundary in the layer with the thickness of 0.5 μm which is half of a grains radius [36]  Fig. 2.15(a). However, the increase in the concentration of $Ca^{2+}$ ions up to 0.2 at.% [36]  Fig. 2.15(b) and even up to 0.5at.% [20] caused accumulation of the $Ca^{2+}$ ions in the layer near the grain boundary. It should be noted that there was not detected the presence of Ca-rich impurity phase on the grain boundary [20,25,36]. The calcium ions concentrated in the thin layer with the thickness of ~5 nm and with the same garnet structure as the grain volume [20,25]. No boundary was visible on HRTEM image between the Ca-rich layer and the grain volume [20]. The same effect has been confirmed for $Mg^{2+}$ ions, where the presence of Mg-rich layer on the grain boundary was reveal for Cr,Ca,Mg:YAG ceramics [25].

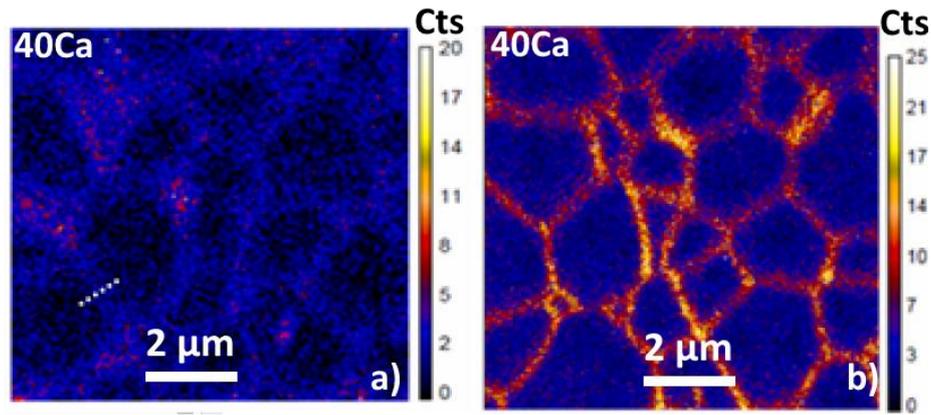

Fig. 2.15: NanoSIMS micrographs of Ca:YAG ceramics showing the spatial distribution of Ca elements for $Ca^{2+}$ concentration a) 0.05 at%, and 0.2 at% (Reproduced with permission from Ref. [36] © Elsevier BV 2023).

Cr,Ca:YAG ceramics after vacuum sintering reveal the presence of Ca-rich layer at the grain boundary with the $Ca^{2+}$ concentration up to 13 at.% which [20] is an order of magnitude more than the solubility limit (0.08% at.% [57]). Based on EDX analysis [20], the incorporation of large concentration of $Ca^{2+}$ ions caused increase in lattice stress which was compensated by formation of $Al_Y$ antisite defects [11]. Also, the high concentration of $Ca^{2+}$ ions was detected at the boundary between YAG grain and $Al_2O_3$ inclusions, while the presence of calcium in aluminum oxide was not detected. Followed air annealing caused decrease the concentration of $Ca^{2+}$ ions in Ca-rich layer at order of magnitude to ~ 1 at.% [20]. Simultaneously, it was detected presence of $Ca^{2+}$ ions in $Al_2O_3$ inclusions and on the ceramics surface ( Fig. 2.16). Air annealing caused segregation part of $Ca^{2+}$ ions from Ca-rich layer into impurities and on the ceramic surface [20,26].

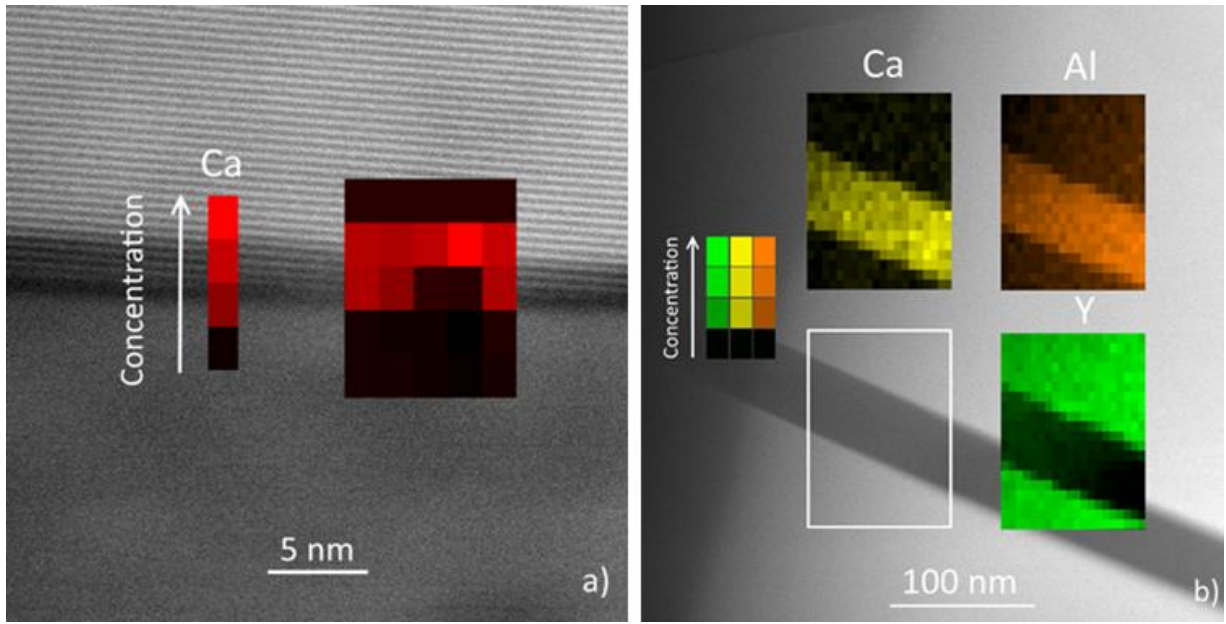

Fig. 2.16: Elemental mapping of Cr,Ca:YAG ceramics of a) Ca at the grain boundary after vacuum sintering and b) $Al_2O_3$ inclusions after air annealing (Reproduced with permission from Ref. [20] © Elsevier BV 2019).

Incorporation of the $Ca^{2+}$ ions in the YAG ceramics occurs thorough formation of $[Ca^{2+}\ldots{}^1\!/_2 V_O]$ charge neutral complex. It was shown formation of charged layers near surface of YAG single crystals enrichment of oxygen vacancies and depletion of cation vacancies that decays with distance typically in the order of several nanometers [75]. This charged layer in single crystal is compensated by the charge from the space charge regions which is the oxygen ions stuck to the surface [75]. Probably the same charged layer creates a net charge on the ceramics grain boundaries. This charged layer supplies the necessary amount of oxygen vacancies in this nanometer thick charged layer to host high concentration of $Ca^{2+}$ ions at the ceramic boundary. Therefore, the concentration of $Ca^{2+}$ ions near the grain boundary is one order of magnitude higher (~1 at.% [20]) than the solubility of $Ca^{2+}$ ions (0.08 at.% [57], or 0.065(15) at.% [36]). Vacuum sintering promotes the formation of oxygen vacancies thus increasing the concentration of $Ca^{2+}$ ions in Ca-rich layer up to 13% at.% [20].

*2.5.2 Distribution of $Mg^{2+}$ ions in $Cr^{4+}$:YAG ceramics.*

The presence of Mg-rich layer in the Cr,Mg:YAG ceramics is unclear now. The Mg-rich layer was detected in the ceramics so-doped with the $Ca^{2+}$ ions, and the presence of the same Mg-rich layer in the ceramics without co-doping with $Ca^{2+}$ ions is under questions (Fig. 2.17). The main reason for accumulation of $Mg^{2+}$ ions near the grain boundary might be the compensation of lattice stress caused incorporations of large $Ca^{2+}$ ions (0.112 nm [56]) instead of small $Y^{3+}$ ions (0.102 nm [56]). In case of Cr,Ca:YAG ceramics, this compensation occurs due to appearance of $Al_Y$ antisite defects [20], while in the case of Cr,Ca,Mg:YAG ceramics [25] this compensation might occurs by segregation of small $Mg^{2+}$ ions (0.089 nm [56]) instead of large $Y^{3+}$ ions (0.102 nm [56]). Moreover, the part of $Mg^{2+}$ ions can be incorporated in YAG ceramics by the interstitial self-compensation mechanism, which is different from the $Ca^{2+}$ ions [11].

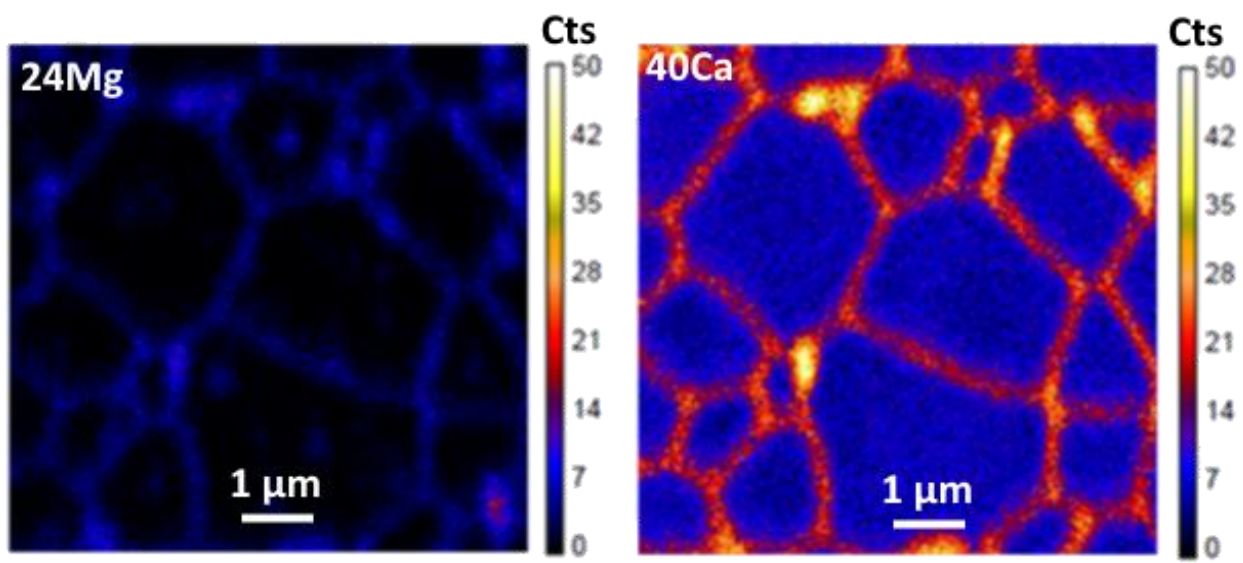

Fig. 2.17: NanoSIMS micrographs of Si,Cr,Ca,Mg:YAG ceramics showing the spatial distribution of Mg and Ca elements (Reproduced with permission from [25] Ref. © Elsevier BV 2021).

*2.5.3 The influence of Ca-rich layer on the $Cr^{4+}$:YAG ceramics properties.*

Ceramics sintered with CaO additives showed smaller grain size in comparison with the pure ceramics [41], indicating the inhibition of grain grow by $Ca^{2+}$ ions. However, the inhibition of grain growth is due to the presence of $Ca^{2+}$ ions in grain volume rather than the formation of Ca-rich layer. This conclusion is based on the reported result ( Fig. 2.18(a)) where the logarithm of grain boundary mobility linearly decreases with increase in the concentration of $Ca^{2+}$ ions beyond the solubility limits (0.065at.%) [36]. It is important that the Ca-rich layer was not detected in the ceramics doped with the $Ca^{2+}$ ions below the solubility limits [36]. The increase in calcium concentration above the solubility limit caused formation of Ca-rich, while the grain boundary migration barely changes.

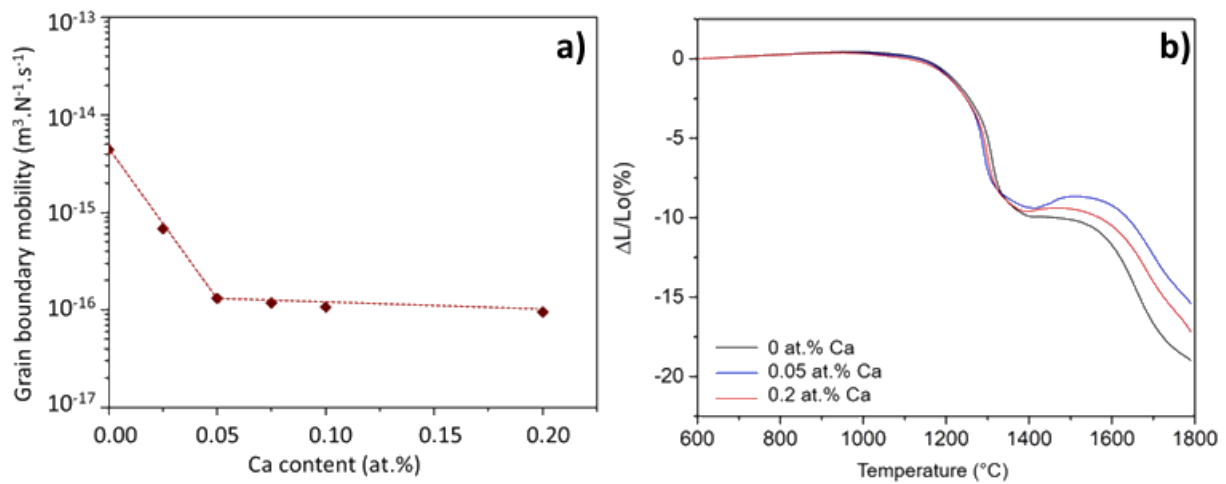

Fig. 2.18: a) Grain boundary mobility and b) densification curves for Ca:YAG ceramics (Reproduced with permission from Ref. [36] © Elsevier BV 2023).

Based on the reported data, it can be concluded that the presence of Ca-rich layer changes the phase transformation rate in YAG ceramics. The phase transformation in YAG ceramics occurs thorough diffusion of $Al^{3+}$ ions thorough YAG/YAP/YAM to $Y_2O_3$ ( Fig. 2.14). Therefore, the presence of Ca-rich layer at the $Al_2O_3$/YAG interface might decrease the $Al^{3+}$ diffusion thus reducing phase transformation rate. This conclusion is based on the reported densification curves for YAG, Ca(0.05at.%):YAG, and Ca(0.2at.%):YAG ceramics ( Fig. 2.18(b)) [36]. The densification curves are the result of sum of volume change due to the phase transformations and the sintering processes. One of the features of densification curves of YAG ceramics is decreasing in the densification at ~1400-1500 °C, due to formation of lighter YAG phase (4.56 g/cm$^3$) from denser YAP phase (5.35 g/cm$^3$) [19]. The increase in Ca-doping content caused a decrease in the densification or even expansion at ~1450 °C indicating the decrease in the phase transformation resulting to shift YAG phase formation to higher temperature. Unfortunately, this assumption was not confirmed experimentally. It should be noted that the change of heating rates had higher impact on the densification [74] than change in Ca-doping content [36] which means that the heating rates should be tuned for the sintering of transparent Ca-doped YAG ceramics.

## 3  $Cr^{3+}→Cr^{4+}$ ions valence transformation in $Cr^{4+}$:YAG ceramics.

The research interest towards $Cr^{4+}$:YAG ceramics base on its application potential, which can be used for tunable lasers or Q-switched pulsed lasers operated at ~ 1 μm. Their applicability potential is based on the unique spectroscopic properties of $Cr^{4+}$ ions in tetrahedral site [10]. Without any other additives, chromium is incorporated in YAG lattice as trivalent ion in octahedral site. $Cr^{3+}↔Cr^{4+}$ ions valence transformation occurs if the ceramics doped with divalent additives such as $Ca^{2+}$ or $Mg^{2+}$ ions [27]. The concentration of $Cr^{4+}$ ions depend on the concentration of additives and the temperature treatment of the ceramics. In this chapter we discussed the $Cr^{3+}↔Cr^{4+}$ ions valence transformation mechanism and the way to achieve high concentration of tetravalent chromium ions in $Cr^{4+}$:YAG ceramics.

### 3.1 The Influence of CaO, MgO, and CaO/MgO additives on the concentration of $Cr^{4+}$ ions in $Cr^{4+}$:YAG ceramics

There are limited data reports the influence of the additives on the $Cr^{4+}$ phase formation in YAG ceramics. A major fraction of the paper uses the combination of CaO and MgO sintering additives in different ratio [14,16,17,22,23,25,26]. Only a few paper reports on the $Cr^{3+} \rightarrow Cr^{4+}$ ions in Cr,Ca:YAG [10,15,22,26,27,67] or Cr,Mg:YAG ceramics [26,52,67]. Now there are insufficient numbers of experimental results to justify the role of additives on the formation of $Cr^{4+}$ ions. The change in the concentrations of the additives and the sintering parameters influenced the $Cr^{3+} \leftrightarrow Cr^{4+}$ ions valence transformation.

The most important parameter of $Cr^{4+}$:YAG ceramics is the absorption at the laser wavelength. Now the existing methods do not allow to calculate the concentration of $Cr^{4+}$ ions with high accuracy [21]. Comparing the different ceramics, the concentration of $Cr^{4+}$ is represented by the absorption of $Cr^{4+}$ ions at 1030 nm. Since $Cr^{3+} \rightarrow Cr^{4+}$ transformation occurs during air annealing, $Cr^{4+}$ absorption spectra can be calculated by the difference between the absorption spectra before and after vacuum sintering. This approach assumes that air annealing does not cause the change in the absorption of the Cr:YAG ceramics except of formation of $Cr^{4+}$ ions. Some papers confirm these assumptions [17,25], while other reports the opposite [23,76].

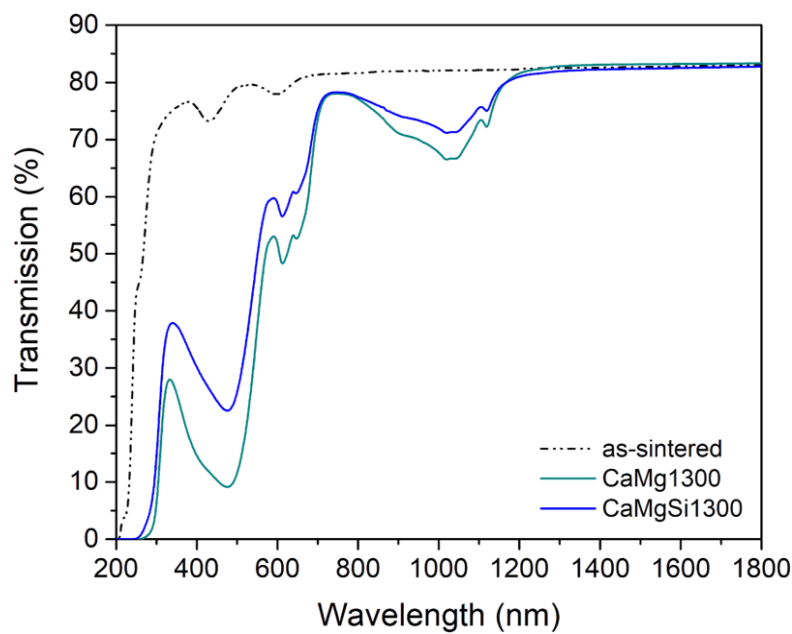

Fig. 3.1: Transmittance spectra for Si,Cr,Ca,Mg:YAG ceramics before (as sintered) and after air annealing (Reproduced with permission from Ref. [25] © Elsevier BV 2021).

The possibility of estimation of $Cr^{4+}$ ions absorption based on assumption that only $Cr^{4+}$ absorption responsible for decrease in the transparency after air annealing. Published results [17,25] and our experience shown that Ca:YAG, Mg:YAG, and Cr:YAG ceramics exhibits the minor changes in the absorption after air annealing. The same results were reported for other $Me^{2+}$-doped ceramics, as in case of Si,Yb,Mg:YAG ceramics even after air annealing at 1450 °C for 40 h [23,77]. However, there is evidence that YAG ceramics doped with divalent additives might change absorption after air annealing due to formation of $[Mg^{2+}…F^{+}]$ or/and $[Ca^{2+}…F^{+}]$ color centers [76]. Moreover, there is paper reports decrease in the transparency of $Cr^{4+}$:YAG ceramics at 1300 nm after air annealing at 1450 °C [23]. Therefore, there is a chance that the estimated absorption of $Cr^{4+}$ ions might be enlarged by the extra absorption form color centers forms during air annealing. It should be noted that this discussion corresponds to the Vis and NIR regions from 400 to ~2000 nm.

*3.1.1 The Influence of CaO additive on the concentration of $Cr^{4+}$ ions in $Cr^{4+}$:YAG ceramics*

The change in the CaO additive has influenced the concentration of $Cr^{4+}$ ions in Cr:YAG ceramics. Cr (0.1at.%),Ca:YAG ceramics sintered at 1750 °C showed change the $Cr^{4+}$ absorption with change in $Ca^{2+}$ ions concentration [27]. The increase in the concentration of CaO additive from 0.04 at.% (Cr/Ca $^{4}/_{1}$) to 0.5 at.% (Cr/Ca $^{1}/_{3}$) caused increase in the absorption of $Cr^{4+}$ ions (1030 nm) from 0.6 to more than 2.2 cm$^{-1}$ [27]. The increase in the concentration of $Cr^{4+}$ ions was more than 3 times. The increase in the concentration of CaO additive % to 0.8 at.% (Cr/Ca $^{1}/_{5}$) caused a decrease in the absorption of $Cr^{4+}$ ions at 1030 nm to 1.8 cm$^{-1}$. It should be noted that the solubility of the $Ca^{2+}$ ions in YAG is lower than 0.08 at.% [57]. Based on the XRD data, it can be concluded that the addition of the CaO additive above the solubility limits reduces the concentration of $Ca^{2+}$ ions in grain volume at half [36]. This indicates that there should be a minor difference in the concentrations of $Ca^{2+}$ ions in the ceramic grains of the ceramics with Cr/Ca ratio $^{4}/_{1}$, $^{1}/_{3}$ and $^{1}/_{5}$. However, the opposite was reported [27].

Such a difference in the concentration of $Cr^{4+}$ ions for the samples with different Cr/Ca ratio can be explained by the influence of the sintering trajectory. The change in the ceramics preparation and the sintering conditions (such as an increase in the sintering temperature to 1800 °C) shown more equal concentrations of $Cr^{4+}$ ions. Cr,Ca:YAG ceramics with the Cr/Ca ratio $^4/_1$, $^1/_1$, $^1/_3$ and $^1/_5$ shown the $Cr^{4+}$ absorption at 1030 nm ~ 1.8(2) $cm^{-1}$, ~ 2(2) $cm^{-1}$, ~ 2.8(2) $cm^{-1}$, and ~ 1.8(2) $cm^{-1}$, respectively. The ceramics with the CaO concentration 0.04 at.% (Cr/Ca ratio $^4/_1$) and with 0.8 at.% (Cr/Ca ratio $^1/_5$) had the same concentration of $Cr^{4+}$ ions. It should be noted that the Ca(0.05at.%):YAG ceramics shown the highest lattice parameters in compare to the other concentrations of CaO additives [36], indicating that the ceramics doped with 0.05 at.% of CaO additives have highest concentration of $Ca^{2+}$ ions in YAG lattice compare to the others concentrations.

Considering the similarities in the $Cr^{4+}$ ions concentrations for Cr,Ca:YAG ceramics with the CaO additives (from 0.04 at.% to 0.8 at.%), it can be concluded that the optimal concentration of CaO is in between 0.05 – 0.08 at.%. Even though the Cr,Ca(0.5at.%):YAG ceramics (Cr/Ca ~ $^1/_3$) have ~ 30% higher concentration of $Cr^{4+}$ ions, exceeding the solubility limits caused formation of Ca-rich layer at the ceramic boundary [36]. After air annealing, a major fraction of $Ca^{2+}$ ions migrate from Ca-rich layer [27] causing deuteriation of the optical properties lead to formation of light scattering centers. This increase in the light scattering can be barely detected due to the increase in the absorption of $Cr^{4+}$ ions.

*3.1.2 The Influence of MgO additive on the concentration of $Cr^{4+}$ ions in $Cr^{4+}$:YAG ceramics*

The main difference between the CaO and MgO additives is the volatility of the latest. Any changes of $Ca^{2+}$ ions in Cr(0.1at.%),Ca(0.5at.%):YAG ceramics was shown [27] even though the CaO concentration exceeds the solubility limits of 0.065 at.% [36] or 0.08 at.% [57]. The same concentration of $Ca^{2+}$ ions was detected at all steps of the ceramic preparations: powder mixture, calcination at 600 °C, and after vacuum sintering at 1750 °C. It should be noted that the ceramics with TEOS sintering additives show decrease in the concentration of $Ca^{2+}$ ions after vacuum sintering at 20% [34], due to the interactions with $SiO_2$. The other case had been found for MgO sintering additive. For example, the sintering of YAG ceramics in wet hydrogen at 1800 °C caused the loss of 80% of $Mg^{2+}$ ions [65].

The evaporation of MgO explains the decrease of concentration of $Cr^{4+}$ ions with increase the sintering temperature for Cr(0.5at.%),Mg(0.5at.%):YAG ceramics [52]. The 1030 nm $Cr^{4+}$ ions absorption for the ceramics sintered at 1750 °C, 1765 °C, 1775 °C, and 1800 °C were ~ 3.6 $cm^{-1}$, ~ 2.9 $cm^{-1}$, ~ 2.5 $cm^{-1}$, and ~ 1.4 $cm^{-1}$, respectively [52]. Reported highest value of $Cr^{4+}$ ions absorption in Cr,Mg:YAG [52,67] is the same as for Cr,Ca:YAG [10,27] and Cr,Ca,Mg:YAG ceramics [16,23]. This indicates that MgO can be efficient in $Cr^{3+} \rightarrow Cr^{4+}$ charge compensation as CaO. However, evaporation of MgO makes using it as single divalent additive more difficult in compared to the CaO. Moreover, $Mg^{2+}$ ions also accumulate in the Mg-rich layer on the ceramic boundary [25], like calcium. Therefore, the better results were obtained using combination of CaO and MgO additives.

*3.1.3 The Influence of CaO/MgO additive on the concentration of $Cr^{4+}$ ions in $Cr^{4+}$:YAG ceramics*

Using the combination of CaO and MgO additives [16,23,25,26] proven to be more successful than the using separately CaO [10,27] or MgO [52,67]. Investigation of the series of Cr(0.1at.%),Me:YAG ceramics (Me – Ca/Mg with ratio $^1/_1$) with the Cr/Me ratio $^1/_1$, $^1/_2$, $^1/_3$, $^1/_4$, $^1/_6$, and $^1/_8$ shown 30% difference in the absorption of $Cr^{4+}$ ions at 1030 nm [16]. The maximum $Cr^{4+}$ ions absorption was found for Cr/Me ratio $^1/_2$ and reached 3.7 $cm^{-1}$. However, the ceramic with the Cr/Me ratio $^1/_1$ exhibits 3.4 $cm^{-1}$ of $Cr^{4+}$ absorption (1030 nm). At this ratio, the Ca concentration was 0.08 at.%, which is exactly as the reported solubility limit in single crystal [57]. Moreover, the transparency of the ceramic with Cr/Me ratio $^1/_1$ was higher than compared to the other ratios [16]. This confirms that the optimal concentration of $Ca^{2+}$ ions should not exceed 0.08. at.%

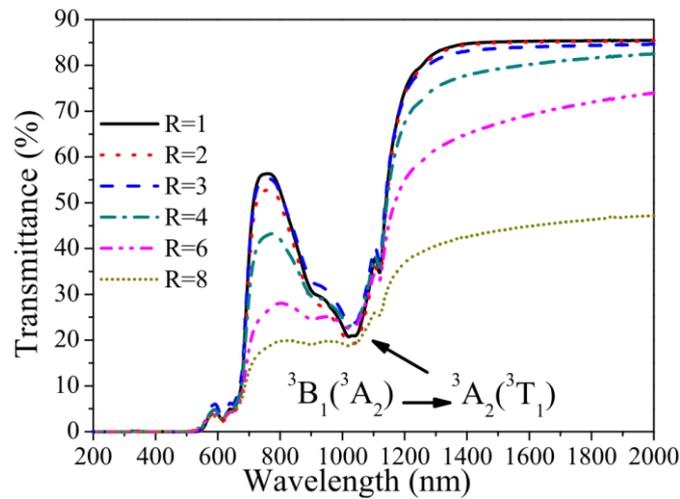

Fig. 3.2: Optical in-line transmittance spectrum of mirror polished Cr,Ca,Mg:YAG ceramics with Cr/Me ratio $^1/_1$ (R=1), $^1/_2$ (R=2), $^1/_3$ (R=3), $^1/_4$ (R=4), $^1/_6$ (R=6), and $^1/_8$ (R=8) (Reproduced with permission from Ref. [16] © Wiley-Blackwell 2015).

The main advantages of using the combination of CaO and MgO is the better optical quality of synthetized samples due to the influence of the MgO additives. However, no positive influence was detected on the concentration of $Cr^{4+}$ ions. The maximum $Cr^{4+}$ ions absorptions (1030 nm) for Cr,Mg:YAG, Cr,Ca:YAG, and Cr,Ca,Mg:YAG ceramics were (3.6 cm$^{-1}$) [52], (3.8 cm$^{-1}$) [20], and (3.7 cm$^{-1}$) [16], respectively. Due to the volatility of MgO additive, CaO remains preferred additives for $Cr^{4+}$ ions formation. In simplified cases, the MgO additive support elimination of pores while CaO conduct $Cr^{3+} \rightarrow Cr^{4+}$ ion valence transformations in $Cr^{4+}$:YAG ceramics. The difference in the role of CaO and MgO indicates that the different approach should be applied for this additive. Probably that the best result can be obtain by weighted precisely of $Cr_2O_3$ and CaO additives to result in $Y_{(1-x)\cdot3}Ca_{3\cdot x}Al_{(1-y)\cdot5}Cr_{5\cdot y}O_{12}$ composition while MgO should be used as the sintering additive.

**3.2 The Influence of $Cr_2O_3$ additive on the concentration of $Cr^{4+}$ ions in $Cr^{4+}$:YAG ceramics**

The change in the concentration of $Cr_2O_3$ additives has influenced the concentration of $Cr^{4+}$ ions. The increase in concentration of $Cr^{3+}$ ions from 0.05 at.% to 0.3 at.% for Cr,Ca,Mg:YAG ceramics sintered in the same conditions lead to the increase of concentration of $Cr^{4+}$ ions in twice [23]. The linear increase in the concentration of $Cr_2O_3$ additive from 0.1 at.% to 1 at.% caused linear increase in $Cr^{4+}$ absorption from ~1.8 cm$^{-1}$ to ~5.4 cm$^{-1}$ for Cr,Mg:YAG ceramics. The reported value for $Cr^{4+}$ ions absorption at 1030 nm for Cr(0.3at.%),Ca,Mg:YAG ceramics (~3.8 cm$^{-1}$) is equal to the highest absorption of $Cr^{4+}$ ions (1030 nm) for Cr(0.1at.%),Me:YAG ceramics (~3.8 cm$^{-1}$) [16]. The $Cr^{4+}$ absorption for Cr(1at.%),Mg:YAG ceramics (~5.4 cm$^{-1}$) is the same as for Cr(0.1at.%),Ca,Mg:YAG ceramics annealed in oxygen (~5.1 cm$^{-1}$).

The same absorption for $Cr^{4+}$:YAG ceramics synthetized with the different type and concentration of divalent additives indicate that the maximum concentration of $Cr^{4+}$ do not limited by the amount of $Cr_2O_3$ additives, at least for chromium concentration equal or larger than 0.1 at.%. It should be noted that the $Cr^{3+}$ ions function as parasitic centers for NIR laser radiations [8,71], therefore the lowest possible concentration of $Cr_2O_3$ additive should be used. The using 0.1 at.% of $Cr_2O_3$ additives allows us to reach highest concentrations of $Cr^{4+}$ ions. However, we do not know what the value of the threshold below of which the concentration of $Cr_2O_3$ additive became the limiting factor. It is important to note that the $Cr_2O_3$ additive is volatile, and the remaining concentration in ceramic after vacuum sintering can be lower than that of calculated. Chemical analysis of Cr(0.1),Ca(0.5):YAG ceramics showed 15% drop in chromium concentration after vacuum sintering [27,34]. The concentration of $Cr^{3+}$ ions can be monitored by characteristic absorption bands in ceramics after vacuum sintering.

*3.2.1 The Influence of the sintering parameters on the concentration of $Cr^{4+}$ ions in $Cr^{4+}$:YAG ceramics*

The amount of $Cr^{4+}$ ions depend on many parameters, such as the concentration of dopants and the parameters of the sintering such as temperature of the vacuum sintering, post annealing, and even from the size of the powders of divalent additives. With the similar parameters of the ceramics preparations, the change in the size of CaO and MgO powders caused change in the concentration of $Cr^{4+}$ ions up to order of magnitude for Cr,Ca,Mg:YAG ceramics [17]. The change in the sintering temperature changes the concentration of $Cr^{4+}$ ions, especially for the ceramics doped with MgO [52,67]. Since the $Cr^{3+} \rightarrow Cr^{4+}$ ion valence transformation occurs during post sintering annealing, the change in the parameters of air annealing changes the concentration of $Cr^{4+}$ ions [10,23]. Until the reaching of saturation point of $Cr^{4+}$ ions concentration, the increase in the temperature and the time of the air annealing exponentially increases the rate of $Cr^{3+} \rightarrow Cr^{4+}$ ion valence transformation [10]. Above the saturation, the $Cr^{4+}$ ions formation occurs but with the lower transformation rate [10]. In other words, $Cr^{3+} \rightarrow Cr^{4+}$ recharge process occurs thorough two different route, the firs are faster and controlled by the diffusion of oxygen thorough the volume of ceramics (see section 3.3) the other one in slower and controlled by the diffusion of the $Ca^{2+}$ ions from the Ca-rich layer to ceramics volume.

Despite all changes in the sintering parameters of $Cr^{4+}$:YAG ceramics, the concentration of $Cr^{4+}$ ions have threshold. So far, the highest reported value of $Cr^{4+}$ ions absorption at 1030 nm is ~ 5 cm$^{-1}$ for both ceramics [23], and single crystals [78]. Now, the concentration of $Cr^{4+}$ ions limit the application of $Cr^{4+}$:YAG material [79]. Despite the change in the concentrations of both divalent dopants and chromium, the $Cr^{4+}$ concentration threshold remains the same [16,20,52]. Moreover, often case that the ceramics doped with higher concentrations of divalent additives exhibit lower concentration of $Cr^{4+}$ ions. This indicates that the used concentration of divalent dopants and chromium is sufficient and other approaches to achieve high concentration of $Cr^{4+}$ ions should be obtained.

The possible limitation factor of $Cr^{4+}$ concentration can be a low solubility of $Me^{2+}$ ions in YAG lattice. The solubility limits of $Ca^{2+}$ ions are ~0.065(15) at.% (to $Y^{3+}$ ions) [36], which was determined from XRD studies and by the appearance of Ca-rich layer. The same Mg-rich layer was detected for Cr(0.1),Ca(0.16),Mg(0.07):YAG ceramics [25] which indicates that the solubility limits of $Mg^{2+}$ ions in YAG lattice is lower than 0.07 at.% (to $Y^{3+}$ ions). This value is closed to the reported earlier solubility limits of $Mg^{2+}$ ions in Mg:YAG ceramics [42]. MgO additives are volatile, thus it is difficult to determine the solubility limits of $Mg^{2+}$ ions. I any case, the solubility limits of $Mg^{2+}$ is comparable or even lower than of the $Ca^{2+}$ ions. The solubility limits of the $Ca^{2+}$ and $Mg^{2+}$ ions are important for calculation of the actual percentage of $Me^{2+}$ ions participated in $Cr^{3+} \rightarrow Cr^{4+}$ ion valence transformation.

Taking into account the solubility limit of $Ca^{2+}$ ions in YAG single crystal (0.08 at.% [57]) and the highest concentration of $Cr^{4+}$ ions [20,23], there is form $^1/_4$ to all of $Ca^{2+}$ ions participate in the $Cr^{3+} \rightarrow Cr^{4+}$ ion valence transformation [19]. Such a broad range is due to the uncertainty in in the $Cr^{4+}$ ions concentrations, which can be calculated with low accuracy [21]. For the ceramics with the 5 $cm^{-1}$ absorption at 1030 nm of $Cr^{4+}$ ions, the concentration of $Cr^{4+}$ ions is in the range from $4 \cdot 10^{18}$ $cm^{-3}$ to $16 \cdot 10^{18}$ $cm^{-3}$, 0.1 at.% of $Cr^{3+}$ ions and 0.08 at.% of $Ca^{2+}$ ions corresponds to $23 \cdot 10^{18}$ $cm^{-3}$ and $11 \cdot 10^{18}$ $cm^{-3}$, respectively. This indicates that the concentration of $Cr^{4+}$ ions can be limited by the solubility limits of $Me^{2+}$ ions. Moreover, even the smaller possible value of $Cr^{4+}$ ions ($4 \cdot 10^{18}$ $cm^{-3}$) indicate that at least 35% of $Ca^{2+}$ ions participate in $Cr^{3+} \rightarrow Cr^{4+}$ ion valence transformation. The present model of chromium valence transformation is based on assumption that the small part of $Ca^{2+}$ ions participate in the $Cr^{3+} \rightarrow Cr^{4+}$. Therefore, the current model should be reconsidered.

**3.3 The proposed model of $Cr^{4+}$ ions formation in Cr:YAG ceramics**

The current understanding of $Cr^{3+} \rightarrow Cr^{4+}$ ions valence transformation based on the assumption that $Cr^{3+}$ and $Ca^{2+}$ ions is in random cation sites after vacuum sintering. This assumption is supported by the experimental data that the $Cr^{4+}$:YAG ceramics after vacuum sintering contain only $Cr^{3+}$. It should be noted that the ceramics after vacuum sintering exhibit small amounts of $Cr^{4+}$ ions (few orders of magnitude lower than after air annealing). Also, it is expected that the ceramics should contain $Cr^{2+}$ ions, however their presence is hard to confirm [80,81]. Chromium transformation into the tetravalent state occurs during air annealing where part of $Cr^{3+}$ ions converted into tetravalent state (Fig. 3.3). The general statement for the $Cr^{4+}$:YAG materials was that a small fraction of $Cr^{3+}$ ions transform into $Cr^{4+}$ [82]. This conclusion arises from the single crystal technology with the difficulties in the estimation of actual concentrations of $Cr^{4+}$ ions.

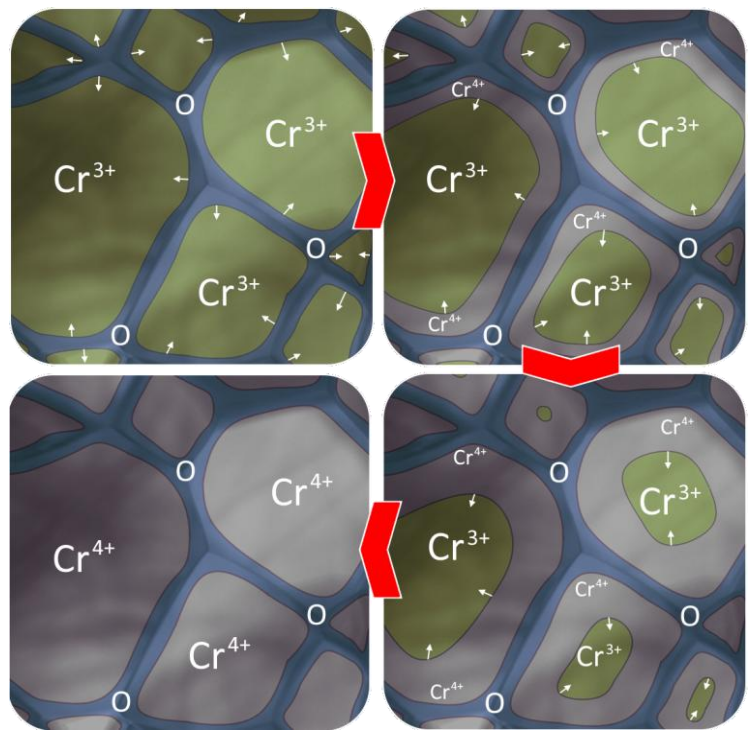

Fig. 3.3: Schematic view of the $Cr^{4+}$:YAG ceramic layer structure during air annealing, where O, $Cr^{3+}$, and $Cr^{4+}$ denote the core, shell, and product layer, respectively: O—oxygen-rich layer at the grain boundary (blue color), $Cr^{4+}$-oxidized layer, where the concentration of $Cr^{4+}$ is constant (violet color), and $Cr^{3+}$-the reduced layer with no $Cr^{4+}$ ions (green color). The arrow indicates the moving direction of the reaction front.

*3.3.1 The traditional view of $Cr^{3+}$→$Cr^{4+}$ ions valence transformation in $Cr^{4+}$:YAG ceramics*

Small concentration of $Cr^{4+}$ ions together with the high concentration of $Me^{2+}$ ions indeed might indicate the random distributions of $Cr^{3+}$ and $Me^{2+}$ ions. The highest concentrations of $Cr^{4+}$ ions were achieved in the ceramics doped with high concentrations of $Me^{2+}$ ions, where the ratio of Cr/Me was larger than 2 [16,27]. This subsection describes the traditional view of $Cr^{3+} \rightarrow Cr^{4+}$ ions valence transformation phenomenon in $Cr^{4+}$:YAG ceramics [10]. The present descriptions are simplified. During the vacuum sintering chromium ions incorporate into YAG lattice in trivalent state in octahedral site [83]. Divalent impurities incorporate as charge neutral complex as $[Me^{2+}\ldots{}^1/_2V_O]$ [20].

During the air annealing the oxygen from the ambient atmosphere incorporates into YAG and can replace the oxygen vacancy from $[Me^{2+}\ldots{}^1/_2V_O]$, with followed recharging $Cr^{3+}$ into tetravalent state thorough formation of $[Me^{2+}\ldots Cr^{4+}]$ complex. This assumption was confirmed by the investigation of the kinetic of the $Cr^{3+} \rightarrow Cr^{4+}$ ions valence transformation, where it was shown that this process in limited by the diffusion of the oxygen thorough ceramics volume [10]. The same conclusion was done from the similar research on $Cr^{4+}$:YAG single crystals [82,84]. The main difference that for single crystals [82,84] and for ceramics [10] were used Wagner and Jander models, respectively (Fig. 3.3). Since $Cr^{3+}$ ions can only occupy octahedral state, $Cr^{3+} \rightarrow Cr^{4+}$ ions valence transformation caused formation of $Cr^{4+}$ ions in octahedral state [82]. Then part of the chromium $Cr^{4+}$ ions from octahedral state migrate into nearest tetrahedral state form $Cr^{4+}$ ions in tetrahedral state.

The weakest point of this model is the transition from $[Me^{2+}\ldots{}^1/_2V_O]$ to $[Me^{2+}\ldots Cr^{4+}]$ complex. This transition should involve migrating of neither chromium nor calcium ions to nearby. This conclusion based on assumption that a major fraction of $Cr^{3+}$ ions and $Ca^{2+}$ ions do not participate in the formation of $[Me^{2+}\ldots Cr^{4+}]$ complexes [82]. In fact, a high percentage of $Ca^{2+}$ ions involve in $Cr^{3+} \rightarrow Cr^{4+}$ ions valence transformation [20]. Taking into account the solubility limits of $Ca^{2+}$ ions in YAG ceramics grain volume (0.065(15) at.% to $Y^{3+}$ [36]) and the concentration of $Cr^{4+}$ ions [20], from $^1/_2$ to all $Ca^{2+}$ ions and from $^1/_7$ to $^1/_4$ of chromium ions is bonded in $[Ca^{2+}\ldots Cr^{4+}]$ charge neutral complexes after air annealing [21]. This disagrees with the statement that a small fraction of chromium and calcium ions participate in $Cr^{4+}$ ions formation.

*3.3.2 A new model of $Cr^{3+}\rightarrow Cr^{4+}$ ions valence transformation in $Cr^{4+}$:YAG ceramics*

The new model of $Cr^{3+}\rightarrow Cr^{4+}$ ions valence transformation is an expansion of the existing one. The main difference is that the new explanation is based on assumption that part of the calcium and chromium ions are incorporates in YAG lattice as $[Ca^{2+}…Cr^{4+}]$ complexes during vacuum sintering. Luminescence studies shown that the Cr,Ca:YAG ceramics after vacuum sintering contain $Cr^{3+}/[Ca^{2+}…V_O]$ charge neutral complexes. In the other words $Ca^{2+}$ ions located near the $Cr^{3+}$ ions in the nearest dodecahedral position. Most probably that the chromium and calcium are incorporates into YAG as $[Ca^{2+}…Cr^{4+}]$ charge neutral complexes during vacuum sintering. The oxygen vacancies destroy these complexes and turn them into $Cr^{3+}/[Ca^{2+}…V_O]$ charge neutral complexes. During the air annealing, the oxygen from ambient atmosphere replaces the oxygen vacancies form $Cr^{3+}/[Ca^{2+}…V_O]$ complexes caused formation of $[Ca^{2+}…Cr^{4+}]$ complexes. According to the proposed model, the concentration of $Cr^{4+}$ ions is determined by the concentration of $Cr^{3+}/[Ca^{2+}…V_O]$ charge neutral complexes after vacuum sintering. The air annealing only activates the destructions of $Cr^{3+}/[Ca^{2+}…V_O]$ and their backward transformation into $[Ca^{2+}…Cr^{4+}]$ complexes.

The main difference between the new model and the accepted one is the distribution of $Ca^{2+}$ ions in YAG grain volume. The accepted model is based on the random distribution of $[Ca^{2+}…V_O]$ and $Cr^{3+}$ ions in the grain volume after vacuum sintering. The new model based on the formation of $Cr^{3+}/[Ca^{2+}…V_O]$ complexes during vacuum sintering. The proposed model based on the earlier described theory of oxidation/reduction-induced chromium ion valence transformations in Cr,Ca:YAG crystals [85]. The $Cr^{3+} \leftrightarrow Cr^{4+}$ ion valence transformation process might differ in ceramics and single crystals. Cr,Ca:YAG crystals growth occurs in oxidizing atmosphere which caused formation of $Cr^{4+}$ ions during the crystal growth. The single crystals synthesized in a reduced environment at lower temperature. Moreover, the single crystal growth occurs from melting, incorporations of both chromium and calcium occurs at the crystallization front. The ceramic synthesis occurs through solid state reactions between initial compounds by mutual diffusions of cations.

# 4  Conclusions

The main motivation to publish this review is to improve the current knowledge related to the sintering of $Cr^{4+}$:YAG ceramics. The main difference in the sintering of $Cr^{4+}$:YAG ceramics compared to the other RE-doped ceramics (Yb:YAG, Nd:YAG, etc.) is necessity to avoid using TEOS sintering additive. Using TEOS support to formation of pore free ceramics and reduce the "methodological" barrier for ceramics technology. The key to obtaining pores-free $Cr^{4+}$:YAG ceramics is the improvement of ceramics preparation technology itself. The presence of CaO, MgO, $Cr_2O_3$, functional additives have a minor influence on the sintering process in compared to the ceramics preparation parameters such as morphology of the initial powders and green body, planetary milling conditions, heating rate and temperature of the isothermal annealing, etc. In the other words, the sintering of $Cr^{4+}$:YAG ceramics faces the same problems as the sintering of Re-doped YAG ceramics without TEOS sintering additive.

The proper tuning of sintering parameters will allow sintering high quality $Cr^{4+}$:YAG ceramics with wide range of concentrations of functional additives. However, using certain concentrations will help to make it easier. The presence of $Cr^{4+}$ ions require the presence of divalent dopants such as CaO or/and MgO. High transparent $Cr^{4+}$:YAG ceramics were synthetized using only CaO or MgO and their combination. The best results were obtained using the combination of CaO and MgO due to the positive influence of MgO on the sintering process and CaO on $Cr^{3+}$ - $Cr^{4+}$ ions valence transformations. The concentration of MgO should be limited by ~ 0.06 wt.%, since extending this value caused the appearance of the second phase. It should be noted that this limit can be affected by evaporation of magnesium. The concentration of CaO should be in between 0.05 – 0.08 at.% (to replace $Y^{3+}$ ions). This concentration is enough to reach ~5 $cm^{-1}$ absorption of $Cr^{4+}$ ions at 1030 nm, which is the highest reported $Cr^{4+}$ ions absorption in YAG ceramics. Using the concentration of CaO higher than 0.08 at.% caused formation of Ca-rich layer at grain boundaries, thus deteriorating the optical properties. The concentration of $Cr^{3+}$ ions should be the smallest possible due to their absorption of laser radiation. The highest concentration of $Cr^{4+}$ ions was achieved using 0.1 at.% of $Cr_2O_3$ additive (to replace $Al^{3+}$ ions).

The new proposed explanations of $Cr^{3+} \leftrightarrow Cr^{4+}$ ion valence transformations highlight the importance of vacuum sintering on $Cr^{4+}$ ions concentration. The new model based on the assumption that the incorporation of calcium and chromium ions in YAG lattice during vacuum sintering occurs in form of $[Ca^{2+}…Cr^{4+}]$ complexes. These complexes then transform into $Cr^{3+}/[Ca^{2+}…V_O]$ due to the interaction with oxygen vacancies. This resulting in the formation of "single" $Cr^{3+}$ ions and $Cr^{3+}/[Ca^{2+}…V_O]$ complexes after vacuum sintering. The air annealing allows to activate $Cr^{3+}/[Ca^{2+}…V_O] \rightarrow [Ca^{2+}…Cr^{4+}]$ transformations, while formation of extra $Cr^{4+}$ ions from "single" $Cr^{3+}$ ions is much slower and have little influence on the overall $Cr^{4+}$ ions concentration. The concentration of $Cr^{4+}$ ions is seeming to be limited by the number of $Cr^{3+}/[Ca^{2+}…V_O]$ complexes which forms during vacuum sintering.

**Appendix**

This chapter explains the article writing style and designation of ceramics. The designation $Cr^{4+}$:YAG indicates that in this case the main characteristic is the presence of $Cr^{4+}$ ions while the presence of chromium ions in different valence state and presence of $Me^{2+}$ additives is unimportant. The designation $Me^{2+}$ means $Ca^{2+}$ or $Mg^{2+}$ ions. The designation Cr,Ca:YAG, Cr,Mg:YAG, or Cr,Ca,Mg:YAG indicates that in this case it is an important type of divalent dopants. The designation such as Cr(0.1),Ca(0.5):YAG is used for highlighted doping concentration of additives, where the concentration of chromium, and divalent dopants was calculated according to replace aluminum, and yttrium ions respectively (in atomic present). The designation Si,Cr(0.1),Ca(0.5):YAG indicates that the ceramics was synthesized with the using of TEOS sintering additives. It should be noted that there are various sources of $Ca^{2+}$ ions can be used such as CaO [24,34,35], $CaCO_3$ [25,36], $Ca(NO_3)_3$ [37], etc. If it is not having influence on the main thought of the text, the CaO denotation will be used for $Ca^{2+}$ ions source. All information from other sources is marked by the link to respected publication/publications. The experimental results without the reference link correspond to unpublished results. More details about these results can be given by the contact with the author.

**Acknowledgements**

This review arises from the knowledge harvested during my doctoral scholarship in Institute for Single Crystals of National Academy of Sciences of Ukraine. The special thanks author would like to say a former supervisor Oleh Vovk and all members of the ceramics laboratory led by prof. dr. Roman Yavetskiy. This work was supported by Polish National Science center, grant: OPUS 23, UMO-2022/45/B/ST5/01487.